%% file: sample-manuscript.tex
\documentclass[manuscript,screen]{acmart}

\AtBeginDocument{%
  }

\setcopyright{acmlicensed}
\copyrightyear{2024}
\acmYear{2024}
\acmDOI{XXXXXXX.XXXXXXX}

\usepackage[utf8]{inputenc}
\usepackage[T1]{fontenc}

\usepackage{graphicx}
\graphicspath{ {./images/} }

\input{addon/pkgloading}
\input{addon/glossary}
\input{addon/glossary-BPM}

\input{addon/macros}
\input{addon/macros-crypto}

\begin{document}

\title{Enabling data confidentiality with public blockchains}


\author{Edoardo Marangone}
\email{marangone@di.uniroma1.it}
\orcid{0000-0002-0565-9168}
\authornotemark[1]
\affiliation{%
	\institution{Sapienza University of Rome}
	\streetaddress{Viale Regina Elena 295, Palazzina E}
	\postcode{00161}
	\city{Rome}
	\country{Italy} 
}

\author{Claudio Di Ciccio}
\orcid{0000-0001-5570-0475}
\affiliation{%
	\institution{University of Utrecht}
	\city{Utrecht}
	\country{The Netherlands}}
\email{c.diciccio@uu.nl}

\author{Daniele Friolo}
\orcid{0000-0003-0836-1735}
\affiliation{%
	\institution{Sapienza University of Rome}
	\city{Rome}
	\country{Italy}}
\email{friolo@di.uniroma1.it}

\author{Eugenio Nerio Nemmi}
\orcid{0000-0001-6518-7863}
\affiliation{%
	\institution{Sapienza University of Rome}
	\city{Rome}
	\country{Italy}}
\email{nemmi@di.uniroma1.it}

\author{Daniele Venturi}
\orcid{0000-0003-2379-8564}
\affiliation{%
	\institution{Sapienza University of Rome}
	\city{Rome}
	\country{Italy}}
\email{venturi@di.uniroma1.it}

\author{Ingo Weber}
\orcid{0000-0002-4833-5921}
\affiliation{%
	\institution{Technical University of Munich, School of CIT \& Fraunhofer}
	\city{Munich}
	\country{Germany}}
\email{ingo.weber@tum.de}

\renewcommand{\shortauthors}{Marangone et al.}

\begin{abstract}
	\input{sections/abstract}
\end{abstract}




\maketitle

\section{Introduction}
\label{sec:intro}
\input{sections/intro}

\section{Example, problem illustration, and requirements}
\label{sec:example}
\input{sections/motivation-and-requirements}

\section{Background}
\label{sec:background}
\input{sections/background}

\input{sections/abe}

\section{The MARTSIA approach}
\label{sec:approach}
\input{sections/approach}
\subsection{Workflow}
\label{sec:approach:workflow}
\input{sections/approach-workflow}
\subsection{Data}
\label{sec:approach:datastructures}
\input{sections/approach-datastructures}

\subsection{Summary and considerations}
\label{sec:approach:considerations}
\input{sections/approach-remarks}

\section{Formal security analysis}
\label{sec:proof}
\input{sections/crypto}

\section{Evaluation}
\label{sec:imptes}
\input{sections/impl-eval}

\section{Related work}
\label{sec:sota}
\input{sections/sota}

\section{Conclusion and future remarks}
\label{sec:conclusion}
\input{sections/conclusion}

\begin{acks}
The work of E.~Marangone, C.~Di~Ciccio, D.~Friolo, E.~N.~Nemmi, and D.~Venturi was partly funded by the SmartDeFi project, Spoke 09, SERICS (PE00000014), under the NRRP MUR program funded by the EU. 
The work of E.~Marangone and C.~Di~Ciccio also received funding from the project PINPOINT (B87G22000450001), under the PRIN MUR program, the Cyber 4.0 projects BRIE and Health-E-Data, and the Sapienza research project ASGARD (RG123188B3F7414A). 
\end{acks}

\bibliographystyle{ACM-Reference-Format}
\bibliography{bibliography.bib}

\end{document}

%% file: addon/pkgloading.tex
\usepackage[english]{babel}
\usepackage{subcaption}
\usepackage{url}
\usepackage{colortbl}
\usepackage{stmaryrd}
\usepackage{amsmath}
\usepackage{mathrsfs}
\usepackage{amsthm}
\usepackage[left]{lineno}
\usepackage{xfrac}
\usepackage{nicefrac}
\usepackage[textsize=scriptsize,backgroundcolor=yellow!40,obeyDraft]{todonotes}
\usepackage[nonumberlist,acronym,sanitize=none]{glossaries}
\glsdisablehyper
\usepackage{comment}
\usepackage[capitalise,nameinlink]{cleveref}
\crefname{algocf}{alg.}{algs.}
\Crefname{algocf}{Algorithm}{Algorithms}
\crefname{definition}{Def.}{Defs.}
\Crefname{definition}{Defition}{Definitions}
\usepackage{tkz-base}
\usetikzlibrary{decorations.pathmorphing,trees,snakes,arrows,shapes,automata,petri}
\usepackage{paralist}
\usepackage{multicol}
\usepackage{booktabs}
\usepackage{enumitem}
\usepackage{multirow}
\usepackage{rotating}
\usepackage{etaremune}
\usepackage{marginnote}
\usepackage{mathtools}
\usepackage{adjustbox}
\usepackage{ifthen}
\usepackage[normalem]{ulem}
\usepackage{lineno}
\usepackage{soul}
\usepackage{listings}
\crefname{lstlisting}{Listing}{Listings}
\usepackage{lipsum}
\usepackage{makecell}
\usepackage{diagbox}
\usepackage[scientific-notation=false]{siunitx}
\usepackage{footmisc}
\usepackage{xspace}
\usepackage{framed}
\usepackage{ifdraft}

%% file: addon/glossary-BPM.tex
\newacronym[\glslongpluralkey={Business Processes}]{bp}{BP}{Business Process}
\newacronym{bpi}{BPI}{Business Process Intelligence}
\newacronym{bpm}{BPM}{Business Process Management}
\newacronym{bpms}{BPMS}{Business Process Management System}
\newacronym{bpmn}{BPMN}{Business Process Model and Notation}
\newacronym{cpn}{CPN}{colored Petri net}
\newacronym{kpi}{KPI}{Key Performance Indicator}
\newacronym{ocbc}{OCBC}{Object-centric Behavioral Constraints}
\newacronym{soa}{SOA}{Service-Oriented Architecture}
\newacronym{pn}{PN}{Petri net}
\newacronym{wf}{WF}{workflow}
\newacronym{wfms}{WfMS}{Workflow Management System}
\newacronym{xes}{XES}{eXtensible Event Stream}
\newacronym{yawl}{YAWL}{Yet Another Workflow Language}
\newglossaryentry{task}{%
	name={task},description={the non-divisible, elementary activity}}

\newglossaryentry{promod}{%
	name={process model},description={the model of a process}
}
\def\LogAlph {\ensuremath{\Sigma}}
\newglossaryentry{logalph}{
	name={log alphabet},description={the process alphabet, as reflected in a log},%
	symbol={\LogAlph}}
\def\Evt {\ensuremath{e}}
\newglossaryentry{evt}{
	name={event},description={a record of an instantaneous fact during the process enactment},%
	symbol={\Evt}}
\def\Trc { \ensuremath{\tau} }

\newglossaryentry{trace}{
	name={trace},description={a sequence of \glsplural{evt}},%
	symbol={\Trc}}
\def\EvtLog {\ensuremath{L}}

\newglossaryentry{evtlog}{
	name={event log},description={a collection of \glstext{evttrace}s},%
	symbol={\EvtLog}}

%% file: addon/macros.tex
\newcolumntype{d}{>{\columncolor{gray!10}}c}
\newcolumntype{m}{>{\columncolor{gray!10}}l}
\newenvironment{iiilist}%
{\begin{inparaenum}[\itshape(i)\upshape]}%
{\end{inparaenum}}

\def\GoodExampleMark{\checkmark}
\def\BadExampleMark{$\times$}

\RequirePackage{xparse}
\NewDocumentEnvironment{AuthNote}{+o+o}{%
	\IfValueT{#2}{\marginnote{\scriptsize{#2}}}%
	\begin{scriptsize}
		\colorbox{gray}%
		{\color{white} Note\IfValueT{#1}{ (#1)}:}%
		\quad%
		\color{brown}
}{%
	\normalcolor
	\end{scriptsize}
}

\RequirePackage{pifont}
\def\CircOne {\ding{192}}
\def\CircTwo {\ding{193}}
\def\CircThree {\ding{194}}

\def\BCircOne {\ding{202}}
\def\BCircTwo {\ding{203}}
\def\BCircThree {\ding{204}}
\def\BCircFour {\ding{205}}

\RequirePackage{lipsum}
\newcommand{\LipsumGray}[1][]{{\color{gray}\ifthenelse{\equal{#1}{}}{\lipsum}{\lipsum[#1]}}}

\RequirePackage{siunitx}
\newcolumntype{D}[1]{S[
	table-omit-exponent,
	round-mode=places,
	round-integer-to-decimal,
	round-precision={#1}]} 

\providecommand{\ie}{{i.e.,}\xspace}
\providecommand{\eg}{{e.g.,}\xspace}

\newcommand{\Req}[1]{\underline{\textbf{R#1}}}

%% file: sections/abstract.tex
Blockchain technology is apt to facilitate the automation of multi-party cooperations among various players in a decentralized setting, especially in cases where trust among participants is limited. Transactions are stored in a ledger, a replica of which is retained by every node of the blockchain network. The operations saved thereby are thus publicly accessible. While this aspect enhances transparency, reliability, and persistence, it hinders the utilization of public blockchains for process automation as it violates typical confidentiality requirements in corporate settings. To overcome this issue, we propose our approach named Multi-Authority Approach to Transaction Systems for Interoperating Applications (MARTSIA). Based on Multi-Authority Attribute-Based Encryption (MA-ABE), MARTSIA enables read-access control over shared data at the level of message parts. User-defined policies determine whether an actor can interpret the publicly stored information or not, depending on the actor's attributes declared by a consortium of certifiers. Still, all nodes in the blockchain network can attest to the publication of the (encrypted) data.
We provide a formal analysis of the security guarantees of MARTSIA, and illustrate the proof-of-concept implementation over multiple blockchain platforms. To demonstrate its interoperability, we showcase its usage in ensemble with a state-of-the-art blockchain-based engine for multi-party process execution, and three real-world decentralized applications in the context of NFT markets, supply chain, and retail. 

%% file: sections/intro.tex
Enterprise applications of blockchain technology are gaining popularity since it enables the design and implementation of business processes involving many parties with little mutual trust, among other benefits~\citep{Weber.etal/BPM2016:UntrustedBusinessProcessMonitoringandExecutionUsingBlockchain,Stiehle22SLR}.
Standard blockchain protocols yield the capability of enabling cooperation between potentially untrusting actors through transparency: Relevant data is made available to all blockchain network participants and can be verified by anyone, thereby removing the need for trust~\citep{Xu.etal/2019:ArchitectureforBlockchainApplications}.
This aspect, in combination with the high-integrity permanence of data and the non-repudiability of transactions offered by 
blockchains, makes them suitable for realizing trustworthy protocols.

However, in multi-party business settings, such as supply-chain or retail processes, 
the involved actors typically have a strong need to keep certain data hidden from some of their competitors, and even more so from most other participants in a blockchain network. 
In fact, fulfilling security and privacy requirements is a key obstacle when it comes to the adoption and implementation of blockchain technology in general~\citep{Privacy1,Privacy2}.
\citet{Corradini.etal/ACMTMIS2022:EngineeringChoreographyBlockchain} and \citet{DBLP:conf/bpm/KopkeN22} emphasize the centrality of these aspects in the context of business process execution on blockchain. 

Simple cryptographic solutions face severe downsides, as discussed in the following.
As \citet{Corradini.etal/ACMTMIS2022:EngineeringChoreographyBlockchain} note, simply encrypting the contents of messages (payload) 
does not guarantee the confidentiality of the information.
Using symmetric encryption requires sharing a decryption key among process participants~\citep{DBLP:journals/fgcs/LuXLWZZ19,Data-Sharing-System-Smart-Cities}, and thus does not allow the sender of the data to selectively control access to different parts of a single message. 
Using asymmetric encryption and encrypting a message with the public key of the recipient requires the sender to create multiple copies of each message (one for each intended reader), which means that the sender can send \textit{different} information to each participant -- i.e., integrity is lost.
Other proposed solutions address the issue via perimeter security: Read access to the (relevant parts of) a blockchain is limited, e.g., by using private blockchain platforms like Hyperledger Fabric with its channels~\citep{Xu.etal/2019:ArchitectureforBlockchainApplications,DBLP:conf/ifip8-1/KopkeB22,Corradini.etal/BCRA2021:ModelDrivenEngineering} or encrypted off-chain storage~\citep{Koepke.etal/BPMBCForum2019:BalancingPrivityEnforceability,Koepke.etal/FGCS2023:DesigningSecureBusiness}.
However, this approach suffers from the same downsides as the use of symmetric encryption above.
Also, permissioned platforms require the presence of trusted actors with the privileged role of managing information exchange \emph{and} the right to be part of the network.
In summary, most of the previous approaches offer ``all-or-nothing'' access: either all participants in some set can access the information in a message, or they receive only private messages and integrity of the data sent to multiple recipients is lost.
In previous work~\citep{Marangone.etal/BPM2022:CAKE}, we introduced a first approach to control data access at a fine-granular level. However, the architecture relied on a central node for forging and managing access keys, thus leading to easily foreseeable security issues in case this single component were to be compromised or Byzantine. 

In this article, we describe MARTSIA (Multi-Authority Approach to Transaction Systems for Interoperating Applications), a fully decentralised approach to secure data sharing in multi-party business cooperation based on public blockchain architectures via \acrfull{maabe}. 
We substantially extend and enhance the early work in which we first presented MARTSIA~\citep{Marangone.etal/EDOC2023:MARTSIA} from conceptual, formal and experimental perspectives. In particular, we provide the following unprecedented contributions:
\begin{iiilist}
	\item We decentralize the certification architecture for user accreditation;
	\item We introduce majority-based mechanisms for system bootstrapping and updates to increase robustness;
	\item We represent the core functionalities offered by our system seen as a protocol and conduct a formal analysis of its security guarantees;
	\item We demonstrate the applicability of our approach in a range of key application domains through the integration with publicly available decentralized applications;
	\item We present a novel implementation on Algorand, showing the cross-platform nature of our solution.
\end{iiilist}

In the following, \cref{sec:example} presents a running example, to which we will refer throughout the paper, and illustrates the problem we tackle. \Cref{sec:background} outlines the fundamental notions upon which our solution is based. In \cref{sec:approach}, we describe our approach in detail. \Cref{sec:proof} analyzes the security of our approach from a formal standpoint. In \cref{sec:imptes}, we present our proof-of-concept implementation and the results of the experiments we conducted therewith. \Cref{sec:sota} reviews related work. \Cref{sec:conclusion} concludes the paper and outlines directions for future work.

%% file: sections/motivation-and-requirements.tex
\begin{figure*}[tb]
	\centering
	\begin{subfigure}[t]{0.67\textwidth}\centering
		\includegraphics[width=\textwidth]{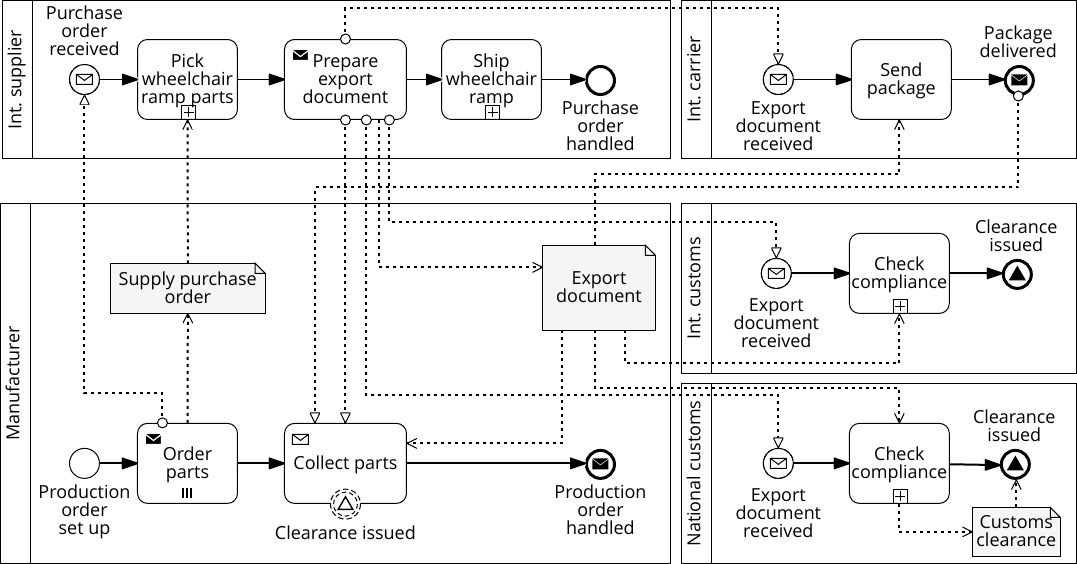}
		\caption[BPMN collaboration diagram]{BPMN collaboration diagram}
		\label{fig:example:bpmn}
	\end{subfigure}%
	\begin{subfigure}[t]{0.35\textwidth}\centering
		\includegraphics[width=0.9\textwidth]{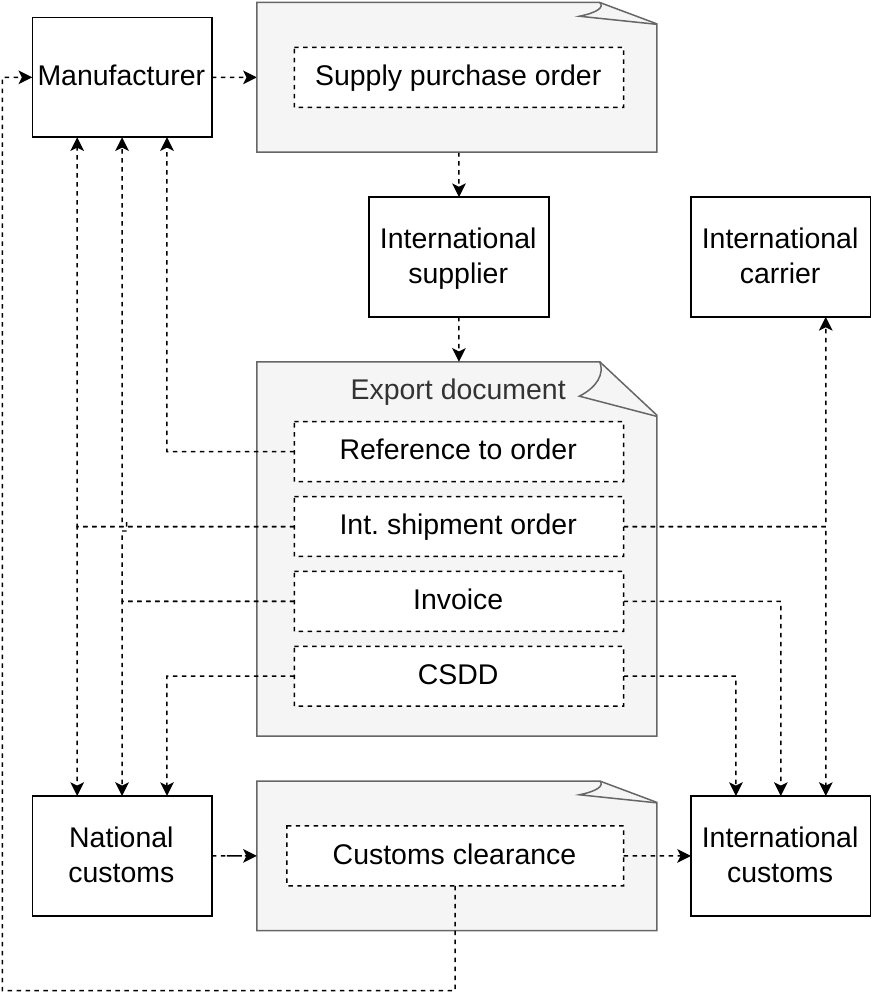}
		\caption[Information Flow Diagram]{Information Flow Diagram}
		\label{fig:example:ifd}
	\end{subfigure}
	\caption{A multi-party process for the assembly of special car parts}\label{fig:example}
\end{figure*}

\Cref{fig:example:bpmn} shows a \gls{bpmn} collaboration diagram~\citep{Dumas.etal/2018:FundamentalsofBPM,DBLP:journals/toit/PourmirzaPDG19} illustrating a fragment of a supply-chain process in the automotive area based on the work of~\citet{Marangone.etal/EDOC2023:MARTSIA}:
the production of a special car for a person with paraplegia. One of the base components for that car is a 
wheelchair ramp to let the person get into the vehicle.
%
A new process instance starts when a \textit{manufacturer} sets up a production order for a modified car. The \textit{manufacturer} orders the necessary parts from an \textit{international supplier} of wheelchair ramps. 
Once all ordered parts have been collected, the supplier prepares the package with the products for delivery. The \textit{national customs} 
verifies the international supplier's \textit{export document} and issues the \textit{customs clearance} once the compliance verification ends with a positive outcome. 
The \textit{international carrier} of the wheelchair ramp supplier checks the documents and delivers the package to the \textit{manufacturer} with an international shipment procedure. The process terminates when the \textit{manufacturer} receives the parts.

%
\begin{table}[tb]
	\caption[An excerpt of the messages exchanged]{An excerpt of the information artifacts exchanged in the process of~\cref{fig:example}}
	\label{tab:messageEncoding:before}
	
    \centering
    \resizebox{0.75\columnwidth}{!}{%
    	\large
		\input{tables/messages-plaintext}	}
\end{table} 
Throughout the paper, we will refer to this scenario as a running example.
In particular, we will focus on the information artifacts marked with a gray background color in \cref{fig:example:bpmn}: 
\begin{inparaenum}[(1)]
	\item\label{item:purchaseorder} the purchase order of the manufacturer; 
	\item\label{item:exportdocument} the export document of the international supplier;
	\item\label{item:customsclearance} the national customs clearance. 
\end{inparaenum}
\Cref{tab:messageEncoding:before} provides example data that they can contain.
\Cref{fig:example:ifd} focuses on their exchange through an Information Flow Diagram (IFD). For one of the artifacts, the IFD also details the parts that it consists of and the intended recipients of those parts: The export document. It encloses multiple records, namely
%
\begin{inparaenum}[({2}.a)]
	\item\label{item:intshipmentorder} the international shipment order, 
	\item\label{item:csdd} the Corporate Sustainability Due Diligence Directive (CSDD), 
	\item\label{item:orderreference} the reference to the order, and
	\item\label{item:invoice} the invoice.
\end{inparaenum}
These records are meant to be accessed by different players.
The shipment order should only be accessible by the international carrier and the two customs bodies,
the CSDD can only be read by the customs authorities,
the order reference is for the manufacturer,
and the invoice is for the manufacturer and the customs bodies.
Differently from the purchase order and the customs clearance (messages \ref{item:purchaseorder}~and~\ref{item:customsclearance}), the four above entries (\ref{item:exportdocument}.\ref{item:intshipmentorder}~to~\ref{item:exportdocument}.\ref{item:invoice}) are joined in a single document for security reasons: separate messages could be intercepted and altered, replaced or forged individually. Once they are all part of a single entity, every involved actor can validate all the pieces of information.
Ideally, in a distributed fashion every node in the network could be summoned to attest to the integrity of that document.
However, the need for separation of indicated recipients demands that only a selected group of readers be able to interpret the parts that are specifically meant for them (see the rightmost column of \cref{tab:messageEncoding:before}).
In other words, though \emph{visible} for validation, the data artifact should not be interpretable by everyone.
The other actors should attest to data encrypted as in \cref{tab:messageEncoding:after}.
This aspect gives rise to one of the requirements 
we discuss next.

\begin{table*}[tb]
	\centering
	\caption[Requirements and corresponding actions in the approach]{Requirements and corresponding actions in the approach}
	\label{tab:requirements}
	\resizebox{1.0\columnwidth}{!}{%
		\normalsize
		\input{tables/requirements}
	}
\end{table*}

\subsection*{Requirements}
%
In recent years, there has been a surge in research on blockchain-based control-flow automation and decision support (see~\cite{Stiehle22SLR} for an overview).
Typically, information shared by actors in a collaborative process is commercial-in-confidence, i.e., shared only with the parties that need access to it, and who are in turn expected to not pass the information on.
Our research complements this work by focusing on secure information exchange among multiple parties in a collaborative though partially untrusted scenario.

\Cref{tab:requirements}
lists the requirements stemming from the motivating use case that drives our approach and a research project in which two authors of this paper were involved.%
\footnote{Cyber 4.0 project BRIE: \url{https://brie.moveax.it/en}. Accessed: October 30, 2024.}
The table highlights the limitations of our past work~\citep{Marangone.etal/BPM2022:CAKE} that we overcome with MARTSIA. Also, it reports on novel features with which we recently developed and were not considered in the first MARTSIA work~\cite{Marangone.etal/EDOC2023:MARTSIA}.
For every requirement, the table indicates the sections in which we discuss the actions taken to meet it.%

Different parties should be granted access to different sections of a confidential information source (\Req{1}, as in the case of the export document in our motivating scenario).
The information source should remain available, immutable, and accountability should be granted for subsequent validations and verifications (\Req{2}, as for the check of the invoice by customs and, more in general, for process mining and auditing~\citep{Klinkmueller.etal/BCForum2019:ExtractingProcessMiningDatafromBlockchainApplications}), without major overheads (\Req{3}) for practical feasibility.
In a distributed scenario such as that of the process in \cref{sec:approach}, where multiple authorities and actors are involved, it is necessary to secure the infrastructure by avoiding that any party can acquire (\Req{4}) or forge (\Req{5}) decryption keys alone.
Our approach should be capable of complementing existing process execution engines to intercept and secure the data flow that characterized multi-party collaborations like the one depicted in \cref{fig:example}~(\Req{6}).
%
Since the system manages sensitive and confidential data, it has to be bootstrapped by a consortium of users who reached a consensus about its core operations (\Req{7}) to avoid rules being changed on the fly by coalitions of malicious players. 
Attempts from malicious users to conduct disruptive operations that could lead to a possible failure or unavailability of the system should be prevented (\Req{8}).
Finally, our system should be platform-independent, so as to be integrable with different secure technologies and do not hinder its adoption in different business contexts (\Req{9}).

Given the requirements above, in the next section we outline the background knowledge our solution is based upon.

%% file: tables/messages-plaintext.tex
\begin{tabular}{|c|c|r l|c|}
\cline{1-5}
   Sender
 & Message
 &
 & Data 
 & Recipients \\
 \cline{1-5}
   Manufacturer
 & \makecell{Supply \\ purchase \\ order \\ (ramp)}
 &
 & 
\begin{lstlisting}[firstline=1,lastline=4]
Company_name:           Alpha
Address:                38, Alpha street
E-mail:                 cpny.alpha@mail.com
Price:                  $5000
\end{lstlisting}
 & \makecell{International supplier}
 \\
 \cline{1-5}   
 & 
 & {\makecell[r]{Int.\\shipment\\order}}
 & 
\begin{lstlisting}[firstline=8,lastline=16]
Manufacturer_company:   Beta
Delivery_address:       82, Beta street
E-mail:                 mnfctr.beta@mail.com
Ramp run:               3
Kickplate:              12
Handrail:               7
Baluster:               30
Guardrail:              25
Amount_paid:            $5000
\end{lstlisting} 
 & \makecell{Manufacturer\\ National customs \\ International customs\\ International carrier}
 \\ 
 \cline{3-5}
   \makecell{International \\ supplier}
 & \makecell{Export \\ document}
 & {CSDD}
 & 
\begin{lstlisting}[firstline=18,lastline=22]
Fundamental_workers_rights:             Ok     
Human_rights:                           Ok
Protection_biodiversity_and_ecosystems: Ok
Protection_water_and_air:               Ok
Combatting_climate_change:              Ok
\end{lstlisting} 
 & \makecell{National customs\\ International customs}
 \\ 
 \cline{3-5} 
 &
 & {\makecell[r]{Reference\\to order}}
 & 
\begin{lstlisting}[firstline=24,lastline=26]
Manufacturer_company:   Beta
Address:                78, Beta street
Order_reference:        26487
\end{lstlisting}
 & \makecell{Manufacturer} 
 \\
 \cline{3-5} 
 & 
 & {Invoice}
 & 
\begin{lstlisting}[firstline=28,lastline=32]
Invoice_ID:             101711
Billing_address:        34, Gamma street
Gross_total:            $5000
Company_VAT:            U12345678
Issue_date:             2022-05-12
\end{lstlisting} 
 & \makecell{Manufacturer\\ National customs\\ International customs}
 \\
 \cline{1-5}
 
 \makecell{Customs \\ clearance}
 & \makecell{National \\ customs}
 &
 & 
\begin{lstlisting}[firstline=36,lastline=40]
Tax_payment:            confirmed
Conformity_check:       passed
Date:                   2022-05-10
Sender:                 Beta
Receiver:               Alpha
\end{lstlisting} 
 & \makecell{Manufacturer\\ International customs} 
 \\
 \cline{1-5}

\end{tabular}

%% file: tables/requirements.tex
	\begin{tabular}{l p{13cm} c c c p{2cm}} \toprule
        &
	& \textbf{CAKE} 
	& \multicolumn{2}{c}{\textbf{MARTSIA}}
        &
	\\\cmidrule{4-5}
	& \textbf{Requirement} & 
        \citeyearpar{Marangone.etal/BPM2022:CAKE}
	&
	\citeyearpar{Marangone.etal/EDOC2023:MARTSIA} &
	This work &
        See
	\\ \midrule
	\Req{1} & Access to parts of messages should be controllable in a fine-grained way based on the attributes that a reader bears, while integrity is ensured 
	& \GoodExampleMark
	& \GoodExampleMark
	& \GoodExampleMark
	& \Cref{sec:approach:datastructures,sec:integration:wfengine}	
	\\
	\Req{2} & Information artifacts should be written in a permanent, tamper-proof and non-repudiable way 
	& \GoodExampleMark
	& \GoodExampleMark
	& \GoodExampleMark
	& \Cref{sec:approach:workflow,sec:integration:wfengine}	
	\\
	\Req{3} & The system should be independently auditable with low overhead 
	& \GoodExampleMark
	& \GoodExampleMark
	& \GoodExampleMark
	& \Cref{sec:approach:workflow,sec:integration:wfengine}	
	\\
	\Req{4} & The decryption key should only be known to the user who requested it 
	& \BadExampleMark
	& \GoodExampleMark
	& \GoodExampleMark
	& \Cref{sec:approach:workflow,sec:integration:wfengine}	
	\\
	\Req{5} & The decryption key should not be generated by a single trusted entity
	& \BadExampleMark
	& \GoodExampleMark
	& \GoodExampleMark
	& \Cref{sec:approach:workflow,sec:integration:wfengine}	
	\\
	\Req{6} & The approach should integrate with control-flow management systems
	& \BadExampleMark
	& \GoodExampleMark
	& \GoodExampleMark
	& \Cref{sec:approach:datastructures,sec:integration:wfengine}
	\\
	\Req{7} & The system should be bootstrapped by a consortium of users
	& \BadExampleMark
	& \BadExampleMark
	& \GoodExampleMark
	& \Cref{sec:approach:workflow,sec:perfomance:analysis}
	\\
	\Req{8} & Parties’ operations should be authorized by a consortium to avoid malicious or disruptive actions
	& \BadExampleMark
	& \BadExampleMark
	& \GoodExampleMark
	& \Cref{sec:approach:workflow,sec:perfomance:analysis}
	\\
	\Req{9} & The system should be platform-independent
	& \BadExampleMark
	& \BadExampleMark
	& \GoodExampleMark
	& \Cref{sec:approach:workflow,sec:swimpl}
	\\
	\bottomrule
\end{tabular}

%% file: sections/background.tex
\DLTs and specifically programmable blockchain platforms serve as the foundation for our work along with \MAABE. In this section, we will explain the basic principles 
underneath these building blocks.

\DLTs are protocols that allow for the storage, processing, and validation of transactions among a network of peers without the need for a central authority or intermediary. These transactions are timestamped and signed cryptographically, relying on asymmetric or public key cryptography with a pair of a private and a public key.
In \DLTs, every user has an account with a unique address, associated with such a key pair. 
Recent \DLT platforms implement additional features such as \emph{multi-signature} (or \emph{multi-sig} for short) \emph{accounts}, namely accounts managed by $N$ entities (each with their own private key) that require at least $n$ signatures, with $1 \le n \le N$, to make the issued transaction valid. 
The shared transaction list forms a ledger that is accessible to all participants in the network. 
A \textbf{blockchain} is a specific type of \DLT in which transactions are strictly ordered, grouped into blocks, and linked together to form a chain.
\DLTs, including blockchains, are resistant to tampering due to the use of cryptographic techniques such as hashing (for the backward linkage of blocks to the previous one), and the distributed validation of transactions.
These measures ensure the integrity and security of the ledger.
Blockchain platforms come endowed with consensus algorithms that allow distributed networks to reach eventual consistency on the content of the ledger~\citep{Nakamoto/2008:Bitcoin:APeer-to-PeerElectronicCashSystem}.
Public blockchains like 
Ethereum~\cite{Wood/2018:Ethereum} and Algorand~\cite{Chen.Micali/TCS2019:Algorand} 
charge fees for the inclusion and processing of transactions. 
These platforms also allow for the use of \textbf{Smart Contracts}, \ie programs that are deployed, stored, and executed on the blockchain~\cite{Dannen/2017:IntroducingEthereumandSolidity,zheng2020overview,10.1145/3564699}.\todo{ACM TOPS added}
Ethereum and Algorand 
support smart contracts with dedicated modules named \EVM and \AVM, respectively.
They are deployed and invoked through transactions, i.e.,
their code is stored on chain and executed by many nodes in the network. 
Outcomes of contract invocations are part of the blockchain consensus, thus verified by the blockchain system and fully traceable.
{\SC}s are at the basis of applications integrating web or standalone front-ends with a blockchain-powered back-end, commonly known as \textbf{\DApps}~\citep{Xu.etal/2019:ArchitectureforBlockchainApplications}.
{\SC}s are typically used to handle the lifecycle and ownership of digital assets that go under the name of tokens.
A well-known class thereof is that of \NFTs~\citep{popescu2021non}.

The execution of \SC code, like transactions, incurs costs measured as \emph{gas} on Ethereum.
In particular, the gas cost is based on the complexity of the computation and the amount of data exchanged and stored. 
Algorand removes the concept of gas fees in favor of a fixed transaction fee structure, having what is called \textit{minimum balance} at its core. The minimum balance is the amount of Algos below which a specific account cannot fall. This value is modified by the deployment of a \SC, the creation of an asset (\eg a token), the opt-in to an asset or the opt-in to a \SC. 
This mechanism is used to prevent spam of unwanted tokens.
Furthermore, the minimum balance can be lowered by opting out or destroying assets and {\SC}s. 
{\SC}s can have global and local variables. Global variables increase the minimum balance in the account of the deployer of the \SC (as that is where these variables are saved). In contrast, local variables increase the minimum balance of the account that opts-in to the \SC because they are saved in that account.

To lower the costs of invoking {\SC}s, external \PtoP systems are often utilized to store large amounts of data~\citep{Xu.etal/2019:ArchitectureforBlockchainApplications}.
One of the enabling technologies is 
\textbf{\IPFS},%
\footnote{\label{foot:ipfs} \href{https://ipfs.tech/}{\nolinkurl{ipfs.tech}}. Accessed: October 30, 2024.}
a distributed system for storing and accessing files that utilizes a \DHT to scatter the stored files across multiple nodes. Like \DLTs, there is no central authority or trusted organization that retains control of all data.
\IPFS uses content-addressing to uniquely identify each file on the network. Data stored on \IPFS is linked to a resource locator through a hash, which--in a typical blockchain integration--is then sent to a smart contract to be stored permanently on the blockchain~\citep{Lopez-Pintado.etal/IS2022:ControlledFlexibilityBlockchainCollaborativeProcesses}. 
In a multi-party collaboration setting like the one presented in \cref{sec:example}, the blockchain 
provides an auditable notarization infrastructure that certifies 
transactions among the participants (e.g., purchase orders or customs clearances). Smart contracts ensure that the workflow is carried out as agreed upon, as described in the works of~\cite{DiCiccio.etal/InfSpektrum2019:BlockchainSupportforCollaborativeBusinessProcesses,Mendling.etal/ACMTMIS2018:BlockchainsforBPM,Weber.etal/BPM2016:UntrustedBusinessProcessMonitoringandExecutionUsingBlockchain}.
Documents like purchase orders, transportation orders, and customs clearances can be stored on \IPFS and linked to transactions that report on their submission. However, data is accessible to all peers on the blockchain. To take advantage of the security and traceability of the blockchain while also controlling access to the stored information, it is necessary to encrypt the data and manage read and write permissions for specific users.

%% file: sections/abe.tex
\begin{sloppypar}
\textbf{\ABE} is a form of public key encryption in which the \emph{ciphertext} (i.e., an encrypted version of a \emph{plaintext} message) and the corresponding decryption key 
are connected through attributes~\cite{ABE,MAABE}. In particular, \CPABE~\cite{CP-ABE,MultiAuthorityCP}
associates each potential user with a set of attributes. Policies are expressed over these attributes using propositional literals that are evaluated based on whether a user possesses a particular property.
In the following, we shall use the teletype font to format attributes and policies.
For example, user \verb|0xB0|\ldots\verb|1AA1| 
is associated with the attributes \texttt{Supplier}, to denote their role, and \texttt{43175279}, to specify the process instance number they are involved in (the \emph{case id}). For the sake of brevity, we omit from the attribute name that the former is a role and the latter a process instance identifier (e.g., \texttt{Supplier} in place of \texttt{RoleIsSupplier} or \texttt{43175279} instead of \texttt{InvolvedInCase43175279}) as we assume it is understandable from the context.
Policies are associated with ciphertexts and expressed as propositional formulae on the attributes (the literals) to determine whether a user is granted access (e.g., \texttt{Carrier or Manufacturer}).

One goal of this work is to move away from a single source of trust (or failure); thus, we consider multi-authority methods.
To decrypt and access the information in a ciphertext, a user requires a dedicated key. 
With \acrfull{maabe}, every authority creates a part of that key, henceforth \emph{\dk}. 
A \dk is a string generated via \MAABE on the basis of
\begin{iiilist}
\item the user attributes, and 
\item a secret key of the authority. 
\end{iiilist}
To generate the secret key (coupled with a public key), the authority requires public parameters composed of a sequence of pairing elements that are derived from a pairing group via Elliptic Curve Cryptography (ECC).
Due to space restrictions, we cannot delve deeper into the notions of pairing groups and pairing elements. We refer to~\cite{MAABE,miller1986use} for further details.
Once the user has obtained a \dk from every required authority, it merges them obtaining the \emph{\fdk} to decrypt the message.
In the Ciphertext-Policy variant of \MAABE, a ciphertext for a given message is generated from the public parameters, the public keys of all the authorities, and a policy.

In our context, users are process participants, messages are the data artifacts exchanged during process execution, ciphertexts are encrypted versions of these artifacts, policies determine which artifacts can be accessed by which users, and keys are the tools given to process parties to try to access the artifacts. In the following sections, we describe how we combine the use of blockchain and Ciphertext-Policy \MAABE to create an access control architecture for data exchanges on the blockchain that meets the requirements listed in \cref{tab:requirements}.
\end{sloppypar}

%% file: sections/approach.tex
In this section, we describe our approach, named Multi-Authority Approach to Transaction System for Interoperating Applications (MARTSIA).
We begin by examining the collaboration among its core software components (\cref{sec:approach:workflow}),
and then illustrate the data structures they handle (\cref{sec:approach:datastructures}).
Finally, we draw considerations on our design choices and security assumptions (\cref{sec:approach:considerations}).

%% file: sections/approach-workflow.tex
\begin{figure*}[tb]
\centering
	\includegraphics[width=\columnwidth]{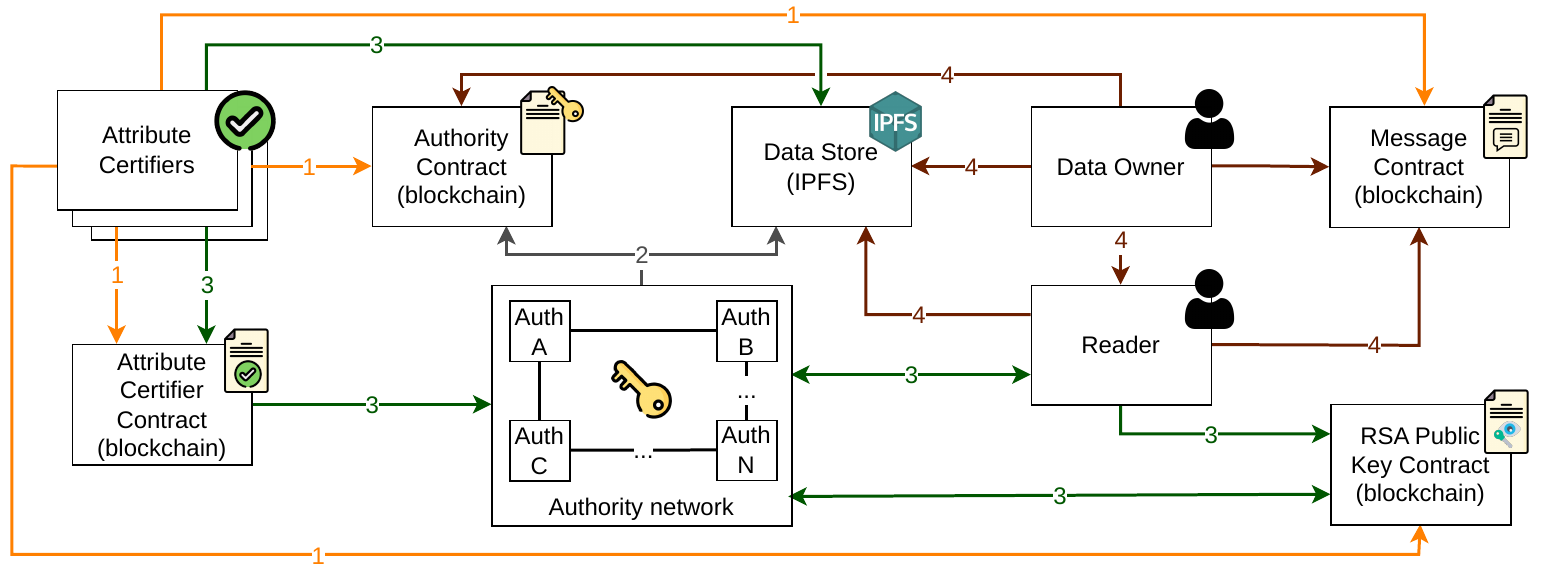}
	\caption{The key components and their interactions in the MARTSIA approach}
	\label{fig:architecture}
\end{figure*}

\Cref{fig:architecture} illustrates the main components of our architecture and their interactions. 
\begin{inparadesc}
	\item[The {\AttCert}] deploys the {\SC}s, assigns the roles to the users of the approach and specifies the attributes characterizing the potential readers of the information artifacts. We assume the {\AttCert}s to hold a blockchain account. Different {\AttCert}s may attest to different pieces of information about potential readers;
	\item[The \DOwner] encrypts the information artifacts (henceforth also collectively referred to as \textit{plaintext}) with a specific access policy. 
    The \DOwner is the sender of the message.
    Considering \cref{fig:example} and \cref{tab:messageEncoding:before}, e.g., the international supplier is the \DOwner of the export document.
    \todo{Was: ``the International Supplier is the \DOwner of the Export document''. We are not using uppercase letters for actors or artifacts. Please check throughout the whole paper.}
    
    \begin{sloppypar}
 	We assume the \DOwner to hold a blockchain account and a Rivest–Shamir–Adleman (RSA)~\citep{DBLP:journals/cacm/RivestSA83} secret/public-key pair.
	\item[The {\Reader}s] 
	are the recipients of the messages. 
	For example, the international supplier is the \Reader of the supply purchase order (see \cref{fig:example,tab:messageEncoding:before}).
    We assume the {\Reader}s to hold a blockchain account, a RSA secret-key/public-key pair, and a global identifier (GID) that uniquely identifies them.    
	\item[The \Auths] calculate separate parts of the secret key for the \Reader (one each). 
	The parties that serve as an \Auth are certified and recognized as such by the {\AttCert}s.
	\item[The \DStore] is a \PtoP repository based on \IPFS. \IPFS saves all exchanged pieces of information in a permanent, tamper-proof manner creating a unique content-based hash as \rloc for each of them.
	\item[The {\SC}s] are used to safely store and make available the {\rloc}s to the ciphertext saved on the \DStore (\textbf{\MeC}), the information about potential readers (\textbf{\AtC}), the data needed by the authorities to generate the public parameters (\textbf{\AutC}), and the RSA public key of the actors (\textbf{\RSAC}). 
	\end{sloppypar}
\end{inparadesc}
%

We divide our approach in four main phases, which we discuss in detail next:
system boot (\cref{sec:workflow:sysboot}), 
initialization (\cref{sec:workflow:init}), 
key management (\cref{sec:workflow:keymgt}), and 
data exchange (\cref{sec:workflow:dataex}).
In the following, the numbering scheme corresponds to the labels in \cref{fig:architecture,fig:workflow:sysboot,fig:workflow:init:keymgt,fig:workflow:dataex}.
Notice that the steps below are designed for public blockchain technologies supporting programmability via smart contracts, regardless of the specific platform in use as per~\Req{9}.

\subsubsection{Step 1: System boot}\label{sec:workflow:sysboot}
%
We assume that all the {\AttCert}s have reached consensus about 
the code of the {\SC}s and the roles of all the users involved (\eg Authorities, {\Reader}s) as described by~\citet{iacr/CerulliCNPS23}.
\begin{figure*}[tb]
	\centering
	\includegraphics[width=\columnwidth]{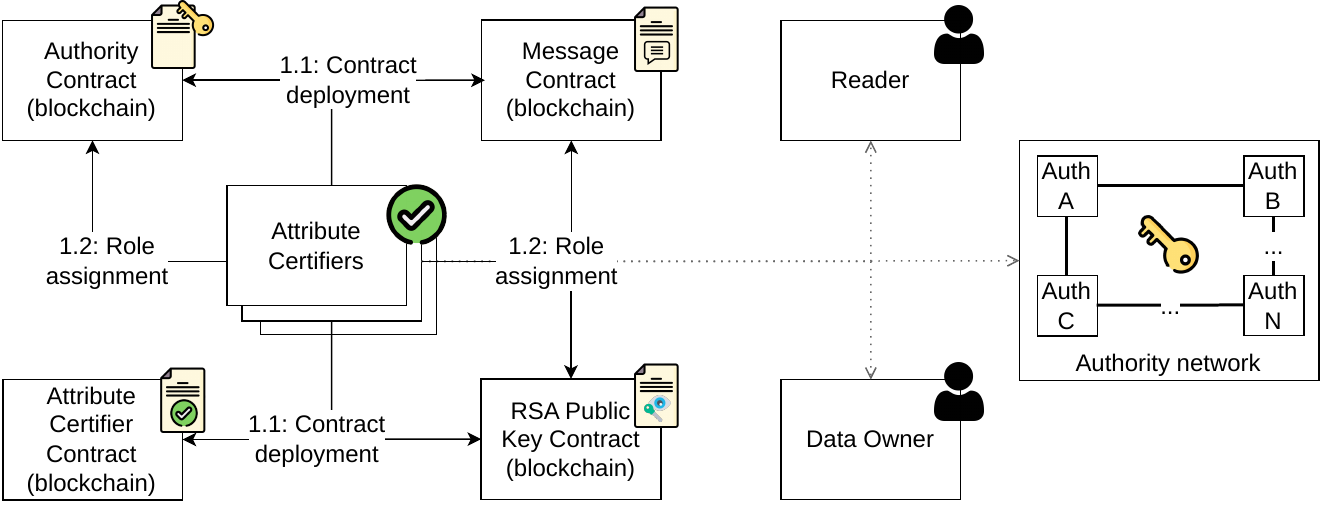}
	\caption{The system boot phase in MARTSIA. Dotted lines are added for clarity to indicate that the roles of \Reader, \DOwner and \Auth are assigned, although no direct message is sent to them by the {\AttCert}s.}
	\label{fig:workflow:sysboot}
\end{figure*}
\begin{inparaenum}[(\bfseries{1}.1\mdseries)]
\Cref{fig:workflow:sysboot} shows the steps that the {\AttCert}s perform to complete the system boot. 
\item\label{step:sysboot:deploy}
First, the {\AttCert}s deploy on the blockchain all the smart contracts needed in our approach. 
To do so, they resort to a majority-based multi-signature account to avoid that all contract instances are determined by a single entity (\Req{7}).%
\footnote{If the blockchain platform in use does not cater for multi-signature accounts, we let one \AttCert call the constructor, though the smart contract remains inactive until the majority of the {\AttCert}s send a transaction to validate and activate it.}
Then, 
\item\label{step:sysboot:roleassign}
the {\AttCert}s assign the roles of \Reader, \DOwner, and \Auth to specific users identified by their blockchain accounts.
The identification of a user based on their role is of utmost importance to mitigate the risk of malicious attempts to assume a false identity, such as impersonating an authority.
To do so, they invoke dedicated methods offered by the \MeC and the \RSAC for the \Reader, by the \MeC for the \DOwner, and by the \AutC for the \Auths.
Notice that similar functions are exposed to let the {\AttCert}s revoke a role later on. 
Also, the role management requires that transactions are sent via a majority-based multi-signature account, thus meeting \Req{8}.
\footnote{If multi-signature accounts are not available, we design the smart-contract methods so that role grants and revocations do not take effect until the majority of the {\AttCert}s invoke them.}
%
\end{inparaenum}%

\subsubsection{Step 2: Initialization} \label{sec:workflow:init}
\begin{figure*}[tb]
	\centering
	\begin{subfigure}[b]{0.27\textwidth}
            \centering
		\includegraphics[width=\columnwidth]{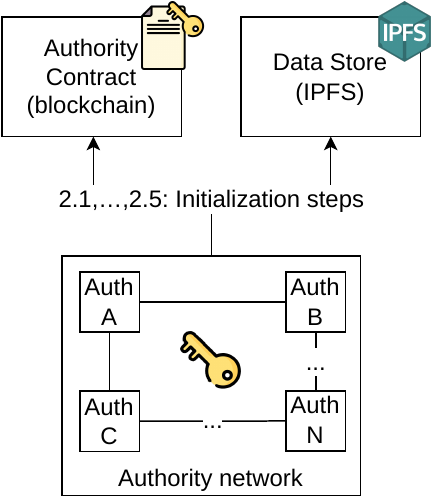}
	\caption{Authority initialization}
	\label{fig:workflow:authinit}
	\end{subfigure}
	\hfill 
	\begin{subfigure}[b]{0.72\textwidth}
            \centering
		\includegraphics[width=\columnwidth]{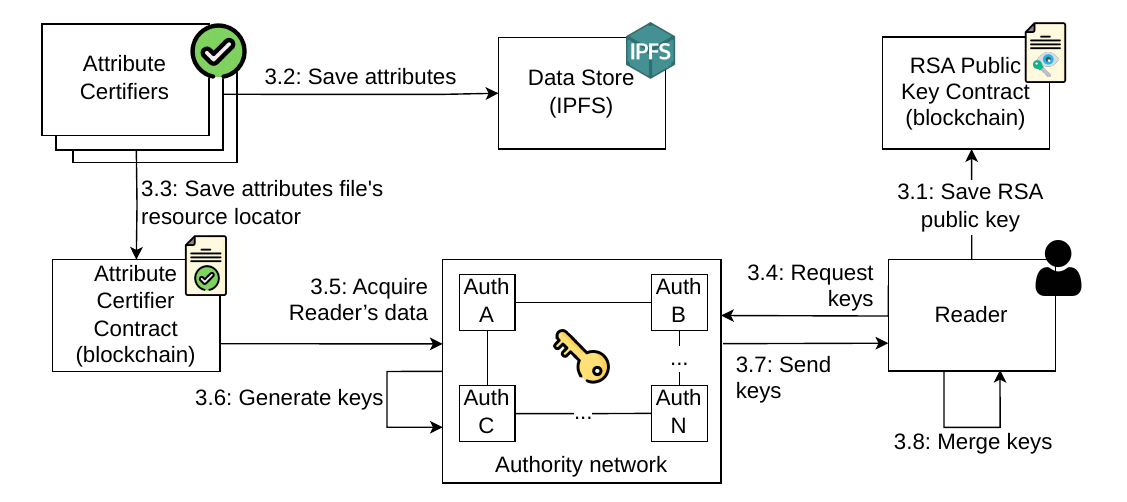}
		\caption{Key management}
		\label{fig:workflow:keymgt}
	\end{subfigure}
	\caption{Authority initialization and key management phases in MARTSIA}\label{fig:workflow:init:keymgt}
\end{figure*}

Here we focus on the network of authorities, as depicted in \cref{fig:workflow:authinit}.
The initialization phase consists of the following five steps.
\begin{inparaenum}[(\bfseries{2}.1\mdseries)]
	\item \label{step:authinit:metadata} 
First, each authority creates a separate file with the metadata of all the authorities involved in the process.%
	\footnote{\label{foot:metadatalink} Notice that metadata are known to all the authorities and all the actors involved in the process.
		Therefore, non-malicious authorities are expected to create an identical file.
		The (same) hash is thus at the basis of the \rloc. 
		As a consequence, anyone can verify whether the authorities behave properly in this step by checking that the {\rloc}s are equal,
		with no need to load the file from the \DStore.}
%
Authorities are responsible for the setting of public parameters that are crucial 
to all the algorithms of \MAABE.
Therefore, we have redesigned the public parameter generation program as a \MPC protocol~\cite{Yao82b,Yao86} to guarantee full decentralization.
More specifically, we adapt a commit-then-open coin-tossing protocol~\cite{Blum81} as follows to generate a random pairing element, that is, the core piece of data described by~\citet{Rouselakis} for \MAABE implementation.
\item 
Each authority posts on the blockchain the hash of a locally generated random pairing element by invoking the \AutC.
\item 
After all the hashes are publicly stored, 
each authority posts the \emph{opening},
namely the previously hashed pairing element in-clear, completing the commit-then-open coin-tossing protocol introduced before. 
\item \label{step:authinit:publicparams} 
Then, every authority 
\begin{iiilist}
	\item verifies that all the hashes of the pairing elements match the respective openings, 
	\item independently combines all posted openings via bitwise XOR, and
	\item uses the output of this operation (the \emph{final shared pairing element}) to calculate the set of public parameters as illustrated by~\cite{Rouselakis}.
\end{iiilist}
\item \label{step:authinit:pkpk} 
Each authority generates its own public-key/secret-key pair by using the authority key generation algorithm of \MAABE. 
\end{inparaenum}
To enable full decentralization and notarization, we resort to the \DStore to save the output of all actions (\textbf{2.\ref{step:authinit:metadata}} to \textbf{2.\ref{step:authinit:pkpk}}) and the \AutC to keep track of the corresponding {\rloc}s.
%


\subsubsection{Step 3: Key management} \label{sec:workflow:keymgt}
The key management phase 
is comprised of the following steps, as illustrated in \cref{fig:workflow:keymgt}:
\begin{inparaenum}[(\bfseries{3}.1\mdseries)]
\item \label{step:key:storersapubkey} 
the {\Reader}s generate a public/private RSA key pair and store the public key on the blockchain. Then,
\item \label{step:key:writeattr} 
the {\AttCert}s save the attributes and the identifying blockchain account addresses of the {\Reader}s on the \DStore and
\item \label{step:key:writeattrlinkonchain} 
the corresponding \rloc on the \AtC so as to make them publicly verifiable on chain.
These operations follow the reaching of a consensus among the {\AttCert}s~\citep{iacr/CerulliCNPS23}.
To this end, the {\AttCert}s operate as a push-inbound oracle~\citep{Basile.etal/BPMBCF2021:BlockchainProcessesDecentralizedOracles}, storing on chain the attributes that determine the role of the \Reader and, optionally, the list of process instances in which they are involved.
For example, the {\AttCert}s store on chain that \verb|0x82|\ldots\verb|1332| is the address of a user that holds the \texttt{Manufacturer} role 
and participates in the process identified by \texttt{43175279}, 
and later that \verb|0xB0|\ldots\verb|1AA1| and \verb|0x9E|\ldots\verb|C885| are {\Reader}s both endowed with the \texttt{Supplier} and \texttt{43175279} attributes, though the former is \texttt{National} and the latter \texttt{International}. 
Whenever a {\Reader} (e.g., the international customs) wants to access the data of a message (e.g., the sections of interest in the Document), they operate as follows:
\item\label{step:key:request}
They request a key to all the authorities, passing the identifying GID as input (we enter the detail of this passage later);
\item 
Each authority seeks the \Reader data (the blockchain address and attributes), and 
obtains them from the \AtC;
\item\label{step:key:gendk}
Equipped with these pieces of information alongside the public parameters, the secret key, and the user's GID, each authority produces a \MAABE decryption key (\dk) 
for the \Reader, and 
\item\label{step:key:return}
sends it back. Once all {\dk}s are gathered from the authorities, 
\item 
the \Reader can merge them to assemble their own \fdk. 
Notice that none of the \Auths can create the \fdk alone (unless specified as such), thus meeting \Req{5}. Also, no user other than the intended \Reader can obtain the key (\Req{4}).
The key management phase can be interleaved with the Data Exchange phase (described below in \cref{sec:workflow:dataex}).
\end{inparaenum}

\paragraph{A note on the privacy of actors and confidentiality of roles}
We remark that {\Reader}s are referred to by their public addresses, thus keeping pseudonimity for the real-world actors. The attribute names are strings that label propositions in policies. We keep them intuitively understandable for the sake of readability here. Obfuscation techniques can prevent external users from guessing the role of {\Reader}s in the process by reading the public policies. Best practices and guidelines on the application of these techniques goes beyond the scope of this paper, though.

\begin{sloppypar}

\paragraph{A note on the security of key requests}
Maliciously obtaining the \fdk of another \Reader is a high security threat.
To tackle this potential issue,
we implement two different communication schemes for {\dk} requests (steps~\textbf{3.\ref{step:key:request}}~and~\textbf{3.\ref{step:key:return}})
since the \fdk can be created only by merging the {\dk}s provided by the {\Auths}:
\begin{inparaenum}[(I)]
	\item\label{list:comm:scheme:ssl}%
	 based on a secure client-server connection, and
	\item\label{list:comm:scheme:blockchain}%
	 mediated by the blockchain. 
\end{inparaenum}
More specifically, we decompose the issue in two challenges, which we describe next alongside the solutions we propose.
First, we want to avoid that information exchanged between a \Reader and an \Auth is intercepted by other parties. To this end, we convey every communication for scheme~\ref{list:comm:scheme:ssl} through a separate client-server \gls{ssl} connection between the \Reader and every \Auth. Within scheme~\ref{list:comm:scheme:blockchain}, 
the \Auth generates the \dk, encrypts it with the RSA public key of the \Reader who requested it, and then returns the encrypted \dk as a payload to a transaction. In this way, only the intended \Reader can decrypt it with their RSA private key.
To avoid any false self-identification as a \Reader through their global identifier (GID), we include a preliminary handshake phase for scheme~\ref{list:comm:scheme:ssl}.
It starts with every \Auth (server) sending a random value (or \textit{challenge}) to the \Reader (client). The latter responds with that value signed with their own RSA private key, so as to let the invoked components verify their identity with the caller's RSA public key stored on chain at step~\textbf{3.\ref{step:key:storersapubkey}}.
This issue is not present in scheme~\ref{list:comm:scheme:blockchain} because the \Reader sends a direct blockchain transaction to the intended \Auth asking for a key. To do that, a \Reader must sign the transaction. Blockchain signatures can be forged only if the signer holds the unique private key associated to the account. Since a {\Reader} is identified by their account address in our approach, counterfeiting their identity is nearly impossible unless the account's private key of the \Reader is stolen. 

An advantage brought by scheme~\ref{list:comm:scheme:blockchain} is that the provision of the \dk by the \Auths to the \Reader is indisputable as their communication is traced on chain (assuming that the \Auths behave honestly).
However, transactions are typically associated to a cost in cryptocurrency.
In contrast, scheme~\ref{list:comm:scheme:ssl} does not incur fees, but traceability and permanence of the exchanged information are not guaranteed.
\end{sloppypar}

\subsubsection{Step 4: Data exchange} \label{sec:workflow:dataex}
\begin{figure}
	\centering
	\includegraphics[width=.9\columnwidth]{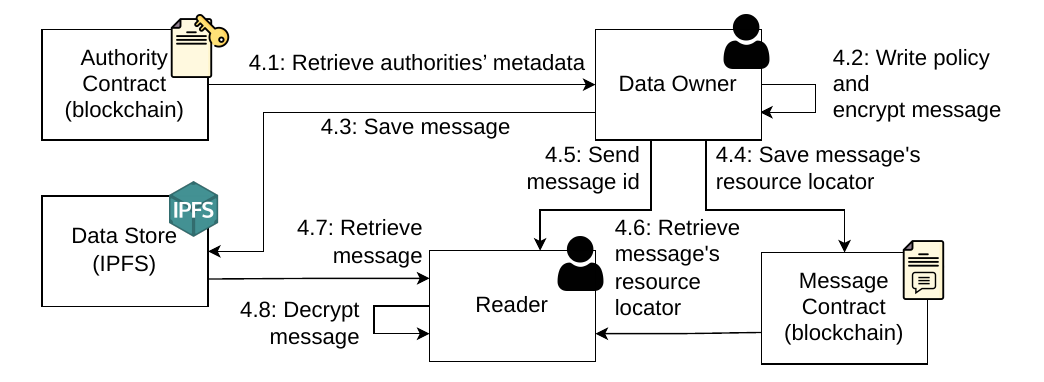}
	\caption{The data exchange phase in MARTSIA}
	\label{fig:workflow:dataex}
\end{figure}

\Cref{fig:workflow:dataex} presents the operations carried out for information storage and access.
As a preliminary operation, the \DOwner verifies that the hash links of the files with metadata and public parameters posted by all the authorities are equal to one another to ascertain their authenticity.\footref{foot:metadatalink}
Then, data is transferred from the \DOwner to the {\Reader}s through the following steps.
\begin{inparaenum}[(\bfseries{4}.1\mdseries)]
\item
\label{step:dataex:retrieveparamspks}%
The \DOwner retrieves the authorities' public keys and public parameters from the blockchain.
\item\label{step:dataex:writepolicy:encrypt}
Then, they write a policy.
Notice that standard \MAABE sets a maximum length for the input files.
In a business process context, this limitation would undermine practical adoption.
To cater for the encryption of arbitrary-size plaintexts, we thus resort to a two-staged hybrid encryption strategy by~\citet{CramerS03}.
First, the \DOwner encrypts via \MAABE a randomly generated symmetric key (of limited size, e.g., \verb|b'3:go+s|\ldots\verb|x2g='|) with 
\begin{iiilist}
	\item the \Auths' public keys,
	\item the policy, and
	\item the public parameters
\end{iiilist}
(obtaining, e.g., \verb|eJytm1|\ldots\verb|eaXV2u|). 
Afterwards, it encrypts the actual information artifact (of any size) via symmetric key encryption scheme~\citep{NIST/FIPS197-2001:AES} using that symmetric key.
In our example scenario, then, the manufacturer does not encrypt via \MAABE the supply purchase order, but the key through which that document is encrypted (and decryptable). 
\item 
Thereupon, the \DOwner (e.g., the manufacturer) saves the encrypted symmetric key and information artifact (plus additional metadata, which we omit here and detail in \cref{sec:approach:datastructures} for the sake of clarity) in one file on the \DStore, 
\item\label{step:dataex:policyenconchain}
sends the file's \rloc to the \MeC, and
\item 
transmits the unique message ID (e.g., \texttt{22063028}) assigned to the file to the \Reader (e.g., the supplier).
As the information artifact is on the \DStore and its \rloc saved on chain, it is written in a permanent, tamper-proof and non-repudiable way, thus meeting requirement \Req{2}.
%
Equipped with their own \fdk,
the \Reader can begin the message decryption procedure.
\item 
At first, the \Reader retrieves the \rloc of the message from the \MeC. 
\item 
Then, once the \Reader obtains the ciphertext from the \DStore, they pass it as input alongside the public parameters (see step \textbf{2.\ref{step:authinit:publicparams}} above) and the \fdk to the \MAABE decryption algorithm running locally. 
\item 
Mirroring the operations explained in step \textbf{4.\ref{step:dataex:retrieveparamspks}}, \MAABE decrypts the symmetric key from the retrieved ciphertext. Notice that only with the symmetric key can the \Reader obtain the original information artifact. 
\end{inparaenum}

\paragraph{A note on the resistance to key counterfeiting}
Since the \DOwner encrypts the symmetric key using also the public keys of the \Auths as one of the parameters (see steps~\textbf{4.\ref{step:dataex:retrieveparamspks}}~and~\textbf{4.\ref{step:dataex:writepolicy:encrypt}}), malicious attackers cannot self-appoint themselves as \Auths and generate working decryption keys.
The attackers, even calculating their own public parameters (step~\textbf{2.\ref{step:authinit:publicparams}}) and thus their own private/public key pairs (step~\textbf{2.\ref{step:authinit:pkpk}}), could not start generating decryption keys that work. The reason is that \Auths use their own private keys to generate a (partial) decryption key \dk (step~\textbf{3.\ref{step:key:gendk}}). Therefore, the would be no correspondence between the forged \dk and a working one since the \Auth's public key used in the encryption phase (steps~ \textbf{4.\ref{step:dataex:retrieveparamspks}}~and~\textbf{4.\ref{step:dataex:writepolicy:encrypt}}) would not match. 
Recall that the \Auths are publicly known (the list is published on chain, see steps~\textbf{1.\ref{step:sysboot:roleassign}}~and~\textbf{2.\ref{step:authinit:metadata}}).

%% file: sections/approach-datastructures.tex
After the analysis of the software components and tasks employed in our approach, we focus on 
messages and policies.

\begin{table*}[tb]
\centering
	\caption[Example of messages stored by MARTSIA upon encryption]{Example of messages stored by MARTSIA upon encryption}
	\label{tab:messageEncoding:after}
	\resizebox{\columnwidth}{!}{%
		\normalsize
		\input{tables/messages-ciphered}
	}
\end{table*}

\subsubsection*{Messages}
\Cref{tab:messageEncoding:before} illustrates the messages we described in our running example in~\cref{sec:example} along with a generated symmetric key for each message. 
\Cref{tab:messageEncoding:after} shows the messages as saved on the \DStore by the \DOwner after the encryption process explained in \cref{sec:approach:workflow} (phase 2).
Each file stored on the \DStore consists of one or more sections to be accessed by different actors (henceforth, \emph{slices}). Every slice is divided in three parts.
\begin{inparadesc}
	\item[The metadata] contain the \emph{message sender} (e.g., \verb|0x82|\ldots\verb|1332| in \cref{tab:messageEncoding:after}), the \emph{case id} (e.g., \texttt{43175279}), and the \emph{message id} that uniquely identifies the message (e.g., \texttt{22063028}).
	\item[The body] is the encrypted information saved as key/value entries (\emph{fields}) for ease of indexation. For security, notice that neither the keys nor the values are in clear. 
	\item[The header] consists of the \emph{encrypted} symmetric \emph{key} generated at step~\textbf{4.\ref{step:dataex:writepolicy:encrypt}}, and the list of field keys that the body contains. In case two or more slices form the message (as in the case of the export document), each is marked with a unique \emph{slice id} (e.g., \texttt{62618638}).  
\end{inparadesc}
We recall that a message is stored on the \DStore and retrievable through a hash, content-based \rloc. The \rloc can thus be attached to process execution data for monitoring and auditability purposes, in compliance with \Req{6}.

%
\subsubsection*{Policies} 
We use policies to specify read grants to message slices, thus enabling fine-grained access control as per \Req{1}. For example, the export document written by the international supplier of process instance \texttt{43175279} is partitioned in four slices as illustrated in \cref{tab:messageEncoding:before}.
We shall use $\mathrm{Attr}$\texttt{@}$X$ as a shorthand notation to indicate that a specific authority $\mathrm{Auth}$ (if $X$ is $\mathrm{Auth}$) or \emph{at least} $n \geq 1$ authorities (if $X$ is $n$\texttt{+}) 
generate the key based on the verification of attribute $\mathrm{Attr}$. 
Compound policies can be formed by joining $\mathrm{Attr}$\texttt{@}$X$ propositions with \texttt{or} and \texttt{and} logical operators.
For instance, \texttt{(Customs@A or Supplier@1+)} declares that only authority \texttt{A} 
can authorize customs, \emph{any} authority can generate the \dk for suppliers,
and only customs or suppliers can read a message. 
We shall thus use the following grammar for policies $P$:
%
\begin{align*}
	P \Coloneqq\: & \mathrm{Attr}\texttt{@}X && \mid \quad  P \texttt{ and } P' &&  \mid \quad  P \texttt{ or } P' \\ 
	X \Coloneqq\: & \mathrm{Auth}            && \mid \quad  n\texttt{+}         &&  
\end{align*}
%

\begin{sloppypar}
Notice that we enable the selection of a specific \dk forger for backward compatibility towards single-authority frameworks.
The downsides are that
\begin{iiilist}
	\item no key is generated if that authority is down (if \texttt{A} crashed, e.g., a user cannot be recognized as a customs body), and
	\item a corrupted authority could take over the generation of an {\fdk} if only one attestation is necessary (theirs).
\end{iiilist}
Therefore, special attention must be paid in the writing of policies.
\texttt{43175279@2+} requires that \emph{at least} two authorities attest to the participation of a user in case \texttt{43175279}. 
%
A user that is not authorized by all the required authorities cannot have 
the \acrlong{fdk} as per the policy. Also, 
whenever multiple authorities are involved in the generation of the \fdk by contributing to a part of it (the \dk),
only the user can compose the \fdk
and decrypt the ciphertext.
\begin{table*}[tb]
\centering
	\caption[Message policy examples]{Message policy examples.}
	\label{tab:messagePolicies}
	\resizebox{\columnwidth}{!}{%
		\huge
		\input{tables/message-policies-new}
	}
\end{table*}
\Cref{tab:messagePolicies} shows the encoding of the policies that restrict access to the slices to specific {\Reader}s based on the attributes that {\AttCert}s attested to in step~\textbf{3.\ref{step:key:writeattr}}. 
For example, the international shipment order is the first slice of the export document. It should be readable by the national and international customs, and by specific actors involved in the process instance: the sender themselves (i.e., the international supplier), the manufacturer, and the international carrier involved in the process instance. Additionally, we exert further constraints on the authorities providing the \dk for specific attributes: customs 
are given the \dk by \Auth \texttt{A}, 
and at least two \Auths must declare that a \Reader is involved in the given process instance.
The other attributes can be attested to by any \Auth.
This composite rule translates to the following expression:
\texttt{Customs@3+ or (43175279@2+ and ((Supplier@1+ and International@1+) or}
\texttt{Manufacturer@1+ or (Carrier@1+ and International@1+)))}.
\end{sloppypar}

%% file: tables/messages-ciphered.tex
\begin{tabular}{|c|l|l|l|}
\hline
   Message
 & Metadata
 & Header
 & Body (slices) \\ \hline
 
   \makecell{Supply \\ purchase \\ order \\ (ramp)}
 & 
\begin{lstlisting}[firstline=1,lastline=3]
sender:         0x82[...]1332
process_instance_id: 43175279
message_id:          22063028
\end{lstlisting}
 & 
\begin{lstlisting}[firstline=2,lastline=4]
EncryptedKey:eJytm1[...]eaXV2u
Fields:       [CV8=[...]6w==, p84W[...]aw==,
               avNL[...]lw==, AOo=[...]mQ==]
\end{lstlisting}
 & 
\begin{lstlisting}[firstline=2,lastline=5]
CV8=[...]6w==: Ruk/[...]QQ==
p84W[...]aw==: 1UTk[...]mA==
avNL[...]lw==: VT+D[...]JQ==
AOo=[...]mQ==: ZUE/[...]6w==
\end{lstlisting} \\ 
 \hline
  
 & 
 & 
\begin{lstlisting}[firstline=8,lastline=14]
SliceId:     62618638
EncryptedKey:eJytWU[...]Bmn4k=
Fields:       [u7o=[...]iw==, TacL[...]IQ==, 
               EhrB[...]lw==, Tiba[...]Wg==, 
               xJ+6[...]SQ==, RPn7[...]Zz==, 
               T3Wq[...]eK==, QXjN[...]LS==, 
               Jmk8[...]fL==]
\end{lstlisting}
 & 
\begin{lstlisting}[firstline=8,lastline=16]
u7o=[...]iw==: py8l[...]mA==
TacL[...]IQ==: MKNH[...]uQ==
EhrB[...]lw==: 5bz0[...]bg==
Tiba[...]Wg==: JahD[...]5Q==
xJ+6[...]SQ==: 3aIm[...]kg==
RPn7[...]Zz==: nRNh[...]GA==
T3Wq[...]eK==: PxRk[...]0g==
QXjN[...]LS==: Lfhk[...]Ug==
Jmk8[...]fL==: Gj43[...]yg==
\end{lstlisting} \\ 
 \cline{3-3}
 \makecell{Export \\ document}
 & 
\begin{lstlisting}[firstline=5,lastline=7]
sender:         0x9E[...]C885
process_instance_id: 43175279
message_id:          15469010
\end{lstlisting}
 & 
\begin{lstlisting}[firstline=16,lastline=20]
SliceId:     19756540
EncryptedKey:eJytm0[...]wtYFo=
Fields:       [ZOSH[...]9B==, BZ8l[...]KY==,
               QeUy[...]WT==, rtdZ[...]hf==,
               8kuG[...]Ts==]
\end{lstlisting}
 & 
\begin{lstlisting}[firstline=17,lastline=21]
ZOSH[...]9B==: sctC[...]nQ==
BZ8l[...]KY==: 5Gln[...]Pg==
QeUy[...]WT==: k49+[...]kA==
rtdZ[...]hf==: lRLd[...]tm==
8kuG[...]Ts==: 25E0[...]uc==
\end{lstlisting} \\ 
 \cline{3-3}
   
 & 
 & 
\begin{lstlisting}[firstline=22,lastline=25]
SliceId:     12191034
EncryptedKey:eJytWU[...]sGlU4=
Fields:       [t8gr[...]QQ==, yuwd[...]vg==,
               1K1d[...]zQ==]
\end{lstlisting} 
 & 
\begin{lstlisting}[firstline=22,lastline=24]
t8gr[...]QQ==: ZJ1v[...]5A==
yuwd[...]vg==: 7XpN[...]4A==
1K1d[...]zQ==: +QbM[...]Cw==
\end{lstlisting} \\ 
 \cline{3-3}
 
 &  
 & 
\begin{lstlisting}[firstline=27,lastline=31]
SliceId:     98546521
EncryptedKey:eJytk9[...]oIJS6=
Fields:       [rjY=[...]KQ==, ZdWC[...]xg==,
               6aLB[...]iw==, VD2h[...]6w==,
               8UmX[...]MQ==]
\end{lstlisting}
 & 
\begin{lstlisting}[firstline=25,lastline=29]
rjY=[...]KQ==: JPAv[...]LA==
ZdWC[...]xg==: w05J[...]Hg==
6aLB[...]iw==: 0wWu[...]vA==
VD2h[...]6w==: eZu7[...]QQ==
8UmX[...]MQ==: sXaB[...]kQ==
\end{lstlisting} \\
 \hline
 
 \makecell{National \\ customs \\ clearance}
 & 
\begin{lstlisting}[firstline=9,lastline=11]
sender:         0x5F[...]FFE1
process_instance_id: 43175279
message_id:          64083548
\end{lstlisting}
 & 
\begin{lstlisting}[firstline=35,lastline=38]
EncryptedKey:eJytWU[...]gGl2Y=
Fields:       [fdoT[...]kA==, 2AkH[...]Rw==,
               0bTn[...]6A==, RZVJ[...]rQ==,
               4TXI[...]zw==]
\end{lstlisting}
 & 
\begin{lstlisting}[firstline=32,lastline=36]
fdoT[...]kA==: fUSZ[...]Bg==
2AkH[...]Rw==: dGp2[...]zA==
0bTn[...]6A==: TuR9[...]bA==
RZVJ[...]rQ==: Pq8U[...]dQ==
4TXI[...]zw==: OIzx[...]Mw==
\end{lstlisting} \\
 \hline
 
\end{tabular}

%% file: tables/message-policies-new.tex
\begin{tabular}{|c|c|c|}
	\hline
	Message                      & Slice                     & Policy                                                                                          \\ \hline
	\makecell[c]{Supply purchase \\ order (ramp)} &                           & \texttt{43175279@2+ and (Manufacturer@1+ or (Supplier@1+ and International@1+))} \\ \hline
	\multirow{4}{*}{\makecell[c]{Export \\ document}} &
	\ref{item:intshipmentorder} &
	\makecell[c]{\texttt{Customs@A or (43175279@2+ and ((Supplier@1+ and International@1+)}\\ \texttt{or Manufacturer@1+ or (Carrier@1+ and International@1+)))}} \\ \cline{2-3} 
	& \ref{item:csdd}           & \texttt{Customs@A or (43175279@2+ and (Supplier@1+ and International@1+))}                      \\ \cline{2-3} 
	& \ref{item:orderreference} & \texttt{43175279@2+ and ((Supplier@2+ and International@1+) or Manufacturer@1+)}                \\ \cline{2-3} 
	& \ref{item:invoice}        & \texttt{Customs@A or (43175279@2+ and ((Supplier@1+ and International@1+) or Manufacturer@1+))} \\ \hline
	\makecell[c]{National customs \\ clearance}   &                            & \makecell[c]{\texttt{Customs@A or (43175279@2+ and Manufacturer@1+)}} \\ \hline
\end{tabular}

%% file: sections/approach-remarks.tex
Thus far, we have described MARTSIA's architecture and the techniques it employs. Here, we draw considerations on the rationale behind our approach and conduct a brief threat analysis. In the following \cref{sec:proof}, we will focus on its security through the lens of formal analysis. 
\paragraph{On the design rationale} \label{note:design-choices}
Our approach is built upon an interplay of components leveraging various techniques. We give a recapitulating description of those, highlighting a few main motives underpinning their conception. More detailed remarks are provided in the form of separate notes and dedicated statements above in this section. 
First of all, the encryption mechanism is based upon Cyphertext-Policy \ABE to allow {\DOwner}s to postulate with a custom policy the attributes identifying who will be authorized to read their data. As a consequence, {\DOwner}s can dictate conditions to restrict access without necessarily knowing the individual authorized users. It follows that keys and attributes have a crucial role. They are thus forged and accredited by networks of multiple {\Auths} (hence the adoption of \MAABE) and {\AttCert}s, respectively, to mitigate the risk of having single points of failure or corruption. Notice that the number of {\Auths} and {\AttCert}s is independent from the {\DOwner}s, {\Reader}s, and especially from ciphertexts, whose amount can rapidly increase during the system's lifetime. Therefore, it can be assumed to be a constant value, which favors system scalability. To reduce communication overhead, we employ a one-time key distribution phase without requiring any further interaction between the {\Reader}s and the \Auths. If a new attribute is to be added or a new \Reader joins the system, the \Auths release the new keys solely to the involved {\Reader}s. Favoring scalability while minimizing communication overhead is also the reason why we employ a distributed \DStore to save primary information pieces like messages, certified attributes, and authority metadata. To let our solution guarantee the properties of auditability and non-repudiation with access-controlled data sharing, the artifacts stored on the \DStore are notarized on a blockchain. In particular, we employ \emph{public} blockchain platforms as their network is composed of an unbounded number of nodes, ensuring stronger data availability and transparency. Anyone inspecting the blockchain can verify that some data was sent through our system, even if in an encrypted fashion. Also, we opt for a \emph{programmable} blockchain to make sure that the information stored complies with our workflow and adheres to consistent data structures.%
\paragraph{On the corruption model} \label{note:corruption-model}
Here, we outline a brief threat analysis to define our setting 
with the corruption model of our approach.
In the initialization step (\cref{sec:workflow:init}), we assume that the majority of the \Auths generating the partial decryption keys (based on the \Reader's attributes) is honest and available, as a consequence of the \MAABE scheme of~\citet{Rouselakis} we employ. 
Our approach employs a consensus protocol like vetKeys~\citep{iacr/CerulliCNPS23} in the system boot and key management steps (\cref{sec:workflow:sysboot,sec:workflow:keymgt}). We thus require that at least half of the involved {\AttCert}s (plus one) be honest.
Finally, we assume that the \DOwner is honest, while we admit a dishonest majority for the {\Reader}s. 

%% file: sections/crypto.tex
In what follows, we formalize MARTSIA as a protocol to discuss its security guarantees. In \cref{subsec:UC} and \cref{subsec:ma-abe} we introduce the Universal Composition framework of \citet{Canetti01} upon which we prove security, and the formal specification of Multi-Authority Attribute-Based Encryption as defined by \citet{Rouselakis}.
In \cref{subsec:protocol} we describe the functionality we aim to implement, the protocol, and state the theorem showing that our protocol realizes such a functionality. Finally, in \cref{subsec:implementation} we argue about the concrete implementation of our protocol.
Before proceeding, we list below a few remarks on the notation we adopt in this section.

\medskip
\noindent
\textbf{Notation.}
We use calligraphic letters to denote sets (\eg $\mathcal{X}$), lower-case letters for variables (\eg $x$), and bold upper-case letters do denote random variables (\eg ${\bf X}$). We write $x \getsr \mathcal{X}$ to indicate that $x$ is picked uniformly at random from $\mathcal{X}$. 
A similar notation is used in the presence of a randomized or probabilistic algorithm $\advA$. Indeed, $x \getsr \advA(\cdot)$ means that $x$ is the output of the randomized algorithm $\advA$. 
All the algorithms we will consider are PPT (Probabilistic Polynomial Time), \ie for any input $x\in\bin^*$, $\advA(x)$ terminates in at most polynomially many steps, in the size of its inputs. We denote with $\secpar \in \NN$ the security parameter
and we assume that all the algorithms take as an input a tuple of arity $\lambda$ whose elements are all $1$, denoted with $1^\secpar$. A function $\nu : \NN \rightarrow [0,1]$ is negligible if, for every polynomial $p(n)$, there exists an $N \in \mathbb{N}$ s.t.\ for each ${n_{0} \geq N}, {\nu\left(n_{0}\right) < \frac{1}{p\left(n_{0}\right)}}$. We denote with  $\negl(\lambda)$ the negligible function in $\lambda$. 
Given two random variables ${\bf X}$ and ${\bf Y}$, we write ${\bf X} \approx_c {\bf Y}$ when ${\bf X}$ and ${\bf Y}$ are computationally close.

\subsection{Universally Composable Multi-Party Computation}\label{subsec:UC}
\input{sections/crypto-UC}

\subsection{Multi-Authority Attribute-Based Encryption}\label{subsec:ma-abe}
\input{sections/crypto-ma-abe}

\subsection{Our protocol}\label{subsec:protocol}
\begin{figure*}
	\centering
	
	\begin{framed}
		\scriptsize
		Runs with the security parameter $1^\secpar$, a data owner $\dataowner$, readers $\readers$, authorities $\auths$, message space $\msgspace$. At the outset, the functionality instantiates a dictionary $\dict$,  an empty set $\attrs$, and an empty map $\certattrs \subseteq \readers\times\attrs$.
		
		\begin{description}[noitemsep]
			\item[Attributes Setup:] On input $(\SetAttr,\attrs',\certattrs')$ from $\attrcert$, if $\attrs$ and $\certattrs$ are empty, set $\attrs=\attrs'$ and $\certattrs=\certattrs'$, and sends $(\SetAttr,\attrs,\certattrs)$ to all parties in $\auths\cup\readers$.
			\item[Message Storage:] On input $(\storemessage,\msg,\pol)$ from $\dataowner$, if $\attrs$ and $\certattrs$ are not empty, set $\dict[\msg]\gets\dict[\msg]\cup\{\pol\}$ and send $(\messagestored,\pol)$ to all the parties.
			\item[Message Read:] On input $(\readmessage,\tilde{\attrs})$ from a reader $\reader\in\readers$,
			check if $(\reader,\attr)\in\mathcal{V}$ for each $\attr\in\tilde{\attrs}$. If the checks pass,
			instantiate a set of messages $\messages\gets\emptyset$ and, for each entry $\pol$ in $\dict$, check if $\tilde{\attrs}$ is an authorized set for $\pol$, and if the check passes, set $\messages\gets\messages\cup\dict[\pol]$.
			Finally, output $(\readmessage,\messages)$ to $\reader$.
			
		\end{description}
	\end{framed}
	\caption{Functionality $\FuncTR$ implemented by our protocol $\pitr$.}
	\label{fig:functr}
\end{figure*}
We first describe the functionality we aim to realize. The functionality $\FuncTR$, described in \cref{fig:functr}, is run by a data owner $\dataowner$, an attribute certifier $\attrcert$ a set of readers $\readers$, and a set of authorities $\auths$.
At the outset, the attribute certifier specifies an attribute space $\attrs$ and a map $\certattrs$ certifying which attributes are assigned to each reader. In particular, each entry will be of the form $(\reader,\attr)$. Then, $\dataowner$ can specify a message $\msg$ and a policy $\pol$ attached to such a message, and query $\FuncTR$ with the command $(\storemessage,\msg,\pol)$. $\FuncTR$ will then store the message, which can be retrieved only by readers owning a set of certified attributes $\tilde{\attrs}$ by querying $\FuncTR$ with $\left(\readmessage,\tilde{\attrs}\right)$, where $\tilde{\attrs}$ is an authorized set for $\pol$.

\noindent{\textbf{Protocol description.}}
\begin{figure*}
	\centering
	
	\begin{framed}
		\scriptsize
		\begin{description}[noitemsep]
			\item[Setup:] All the authorities in $\auths$ engage in $\Func_\setup$. At the end of the computation, each authority will receive the public parameters $\params$. Then, each $\auth\in\auths$ computes $(\pk_\auth,\sk_\auth)\getsr\authsetup(\params)$ and sends $(\params,\pk_\auth)$ to each party \footnotemark.
			Then, the attribute certifier $\attrcert$, on input a set $\attrs'$ of certified attributes and a map $\certattrs'\subseteq \readers \times\attrs$,
			generates the smart contract $\FuncSC$ with authorities $\auths$, readers $\readers$, data owner $\dataowner$, message space $\msgspace=\bin^*$
			and makes a query to $\FuncSC$ with $(\SetAttr,\attrs',\certattrs')$. 
			
			\item[Key generation:] 
			For each $\reader\in\readers$ and each pair $(\reader,\attr)\in\certattrs$, the authority $\auth=\polpred(\attr)$ computes $\dk_{\reader,\attr}\getsr\kgen(\params,\reader,\sk_{\auth},\attr)$, and handles $\dk_{\reader,\attr}$ to $\reader$.
			\item[Encryption:] 
			$\dataowner$, on input $\msg$ and $\pol$\footnotemark, computes $\cipher\getsr\formalMAABE.\enc(\params,\msg,\pol,\{\pk_\auth\}_{\auth\in\auths})$. Then, queries $\FuncSC$ with $(\SendCtx,\cipher,\pol)$.
			\item[Decryption:] When a reader $\reader\in\readers$ wants to retrieve messages w.r.t. a set of attributes $\tilde{\attrs}$ for which he owns the corresponding decryption keys $\{\dk_{\reader,\attr}\}_{\attr\in\tilde{\attrs}}$, queries $\FuncSC$ with $(\RetrieveCtx,\tilde{\attrs})$, and receives $(\RetrieveCtx,\ctxs)$ from $\FuncSC$. Then,
			for each $\cipher\in\ctxs$, output the corresponding $\msg=\formalMAABE.\Dec(\params,\cipher,\{\dk_{\reader,\attr}\}_{\attr\in\tilde{\attrs}})$.
		\end{description}
	\end{framed}
	\caption{Our protocol $\pitr$ realizing $\FuncTR$.}
	\label{fig:pitr}
\end{figure*}
Our protocol $\pitr$ is described in \cref{fig:pitr}. We investigate the security of $\pitr$ in the $(\Func_\setup,\FuncSC)$-hybrid model. We use as a main ingredient a multi-authority ABE scheme $\formalMAABE=(\formalMAABE.\setup,\formalMAABE.\authsetup,\allowbreak\formalMAABE.\kgen,\allowbreak\formalMAABE.\enc,\allowbreak\formalMAABE.\dec)$. Since we aim for full decentralization, we require that the setup algorithm of the MA-ABE scheme is run with a multi-party computation protocol by all of the authorities. Since thanks to \citet{Yao82b}, we are aware of general-purpose multi-party computation protocols (i.e., for any function), we can assume the existence of a functionality $\Func_\setup$ implementing $\formalMAABE.\setup(1^\secpar)$. 
\begin{figure}
	\centering
	\begin{framed}
		\scriptsize
		Runs with authorities $\auths$, readers $\readers$, a data owner $\dataowner$, 
		an attribute certifier $\attrcert$, message space $\msgspace$.
		Instantiates empty variables $\attrs$, $\certattrs$, and an empty dictionary $\dict$ to store access policies with the corresponding ciphertexts. Additionally, runs with the following commands:
		\begin{description}[noitemsep]
			\item[Attributes Set:] On input $(\SetAttr,\attrs',\certattrs')$ from $\attrcert$, if $\attrs$ and $\certattrs$ are empty, set $\attrs=\attrs'$ and $\certattrs=\certattrs'$, and send $(\SetAttr,\attrs,\certattrs)$ to all parties.
			\item[Ciphertext Storage:] On input $(\SendCtx,\cipher,\pol)$ from $\dataowner$, set $\dict[\pol]\gets\dict[\pol]\cup\{\cipher\}$.
			\item[Ciphertext Retrieval:] On input $(\RetrieveCtx,\tilde{\attrs})$ from a reader $\reader\in\readers$,
			instantiate a a set $\ctxs\gets\emptyset$. Then,
			for each entry $\pol$ in $\dict$ such that $\tilde{\attrs}$ is an authorized set of $\pol$ set $\ctxs\gets\ctxs\cup\dict[\pol]$.
			Finally, output $(\RetrieveCtx,\ctxs)$ to $\reader$.
			
			\item[Dictionary Retrieval:] On input $(\RetrieveDict)$ from any party $\party$, output $(\RetrieveDict,\dict)$ to $\party$. 
		\end{description}
	\end{framed}
	\caption{Public Smart Contract Functionality $\FuncSC$.}
	\label{fig:funcsc}
\end{figure}
To simplify the protocol description, we also assume the existence of a public smart contract functionality $\FuncSC$\footnote{The idea of modeling a smart contract as a UC functionality was already adopted by \citet{BaumDD20}.}, described in \cref{fig:funcsc} in which $\attrcert$ first stores the attributes $\attrs$ and the map $\certattrs$ (by querying $\FuncSC$ with $(\SetAttr,\attrs,\certattrs)$). Then $\dataowner$ stores encrypted messages $\cipher$ w.r.t. a policy $\pol$ (by querying $\FuncSC$ with $(\SendCtx,\cipher,\pol)$) that will be retrieved by an authorized reader $\reader$  by querying $\FuncSC$ with $(\RetrieveCtx,\tilde{\attrs})$ for a set of certified attributes $\tilde{\attrs}$ and then decrypted by $\reader$ with its decryption keys previously obtained for each attribute in $\tilde{\attrs}$. 
Since $\FuncSC$ is a public smart contract, we further equipped $\FuncSC$ with a command $(\RetrieveDict)$ that can be queried by any party to retrieve the data structure containing all the encrypted messages.

The underlying multi-authority ABE scheme $\formalMAABE$ is instantiated with attribute universe $\attrs$, authority universe $\auths$, global identities  $\idents=\readers$, and message space $\msgspace=\bin^*$. 
The theorem below 
claims that $\pitr$ realizes $\FuncTR$ by assuming a corrupted minority of authorities (i.e. $\left\lceil\frac{|\auths|}{2}\right\rceil-1)$, a majority of corrupted readers $\readers$,
a corrupted attribute certifier $\attrcert$, and an honest data owner $\dataowner$.
\begin{sloppypar}
\begin{theorem} \label{thm:protocol}
	
	Let $\formalMAABE$ be a Multi-Authority Attribute-Based Encryption Scheme with message space $\bin^*$, authorities $\auths$, global identities $\readers$ enjoying correctness (\cref{def:maabe}) and CPA-Security (\cref{def:maabe-cpa}). Let 
	$\parties_1=\auths$, $\parties_2=\readers$, $\parties_3=\{AC\}$ and $\parties_4=\{\dataowner\}$.
	Then the protocol $\pi_\mathsf{TR}$ described above $\left(\left\lceil\frac{n_1}{2}\right\rceil-1,n_2-1,n_3,0\right)$-securely realize $\FuncTR$ in the $(\Func_\setup\FuncSC)$-hybrid model (\cref{def:mult-parties-uc}).
\end{theorem}
\end{sloppypar}
\input{sections/crypto-proofUC}
\footnotetext[3]{The $\params$ value of the corrupted authorities may differ. Parties will take the parameters having the same value from the majority of the authorities.}
\footnotetext[4]{We require 
	that $\pol$ authorizes attributes 
	of the form ${\{\attr_\auth=(\attr',\auth)\}_{\auth\in\mathcal{L}}}$, where $\mathcal{L}$ is a subset of $\auths$ of cardinality $\left\lfloor\frac{|\auths|}{2}\right\rfloor+1$.}

\begin{remark}
	Our protocol is proven in a UC model variant requiring static security (i.e., the corrupted parties are fixed at the beginning of the protocol). In our functionality $\FuncTR$, the attribute certifier may assign attributes to the readers only once at the beginning. 
	Even though this variant of $\FuncTR$ is sufficient for the MARTSIA setting, we 
	we may possibly augment $\FuncTR$ by allowing $\attrcert$ to update the set $\certattrs$ at any time and then proving security of a variant of our protocol in the adaptive UC model (i.e., where the adversary may adaptively corrupt parties during the protocol execution) by relying on the recent fully-adaptive multi-authority attribute-based encryption schemes of \citet{DattaKW23}. In this protocol variant, when $\attrcert$ adds new values of the type $(\reader,\attr)$ in $\certattrs$, the authorities are allowed to release the new keys to the respective readers independently of when $\dataowner$ sends ciphertexts to $\FuncSC$.
\end{remark}
\subsection{Protocol implementation}\label{subsec:implementation}
Our concrete implementation of $\pitr$ is realized by using the multi-authority ABE implementation of \citet{Rouselakis} for linear access policies. We recall that the global paremeters $\params$ of \citet{Rouselakis} are of the form $\left(p,\mathbb{G},g,H,F,\attrs',\bar{\attrs}\right)$, where $\attrs=\attrs\cup\bar{\attrs}$ are the attributes chosen by $\attrcert$, $H$ and $F$ are publicly known functions, $\mathbb{G}$ is a publicly known group of prime order $p$, and $g$ is a random generator for $\mathbb{G}$. 
Notice that the only value that can negatively affect the protocol's security $g$ should be randomly chosen. Therefore, instead of using a general purpose MPC protocol for the $\setup$ algorithm, we implemented $\Func_\setup$ via a simple on-chain commit-and-open multi-party coin tossing protocol outputting a random value $g$~\citep{AndrychowiczDMM14,BottaFVV21}.
We implemented $\FuncSC$ with both Ethereum and Algorand smart contracts. We stress that $\FuncSC$ can also be implemented with Bitcoin by relying on digital signatures to authenticate the transactions made by the authorities, the data owner, the attribute certifier, and the readers by requiring all the players to post a certified verification key to the ledger at the beginning of the protocol execution.

For the sake of simplicity, in our formal protocol description of $\pitr$ we made the authorities handle the decryption keys of $\formalMAABE$ to the authorized readers via an off-chain communication through a secure channel. In our actual implementation, we have also provided the option to use the ledger as a communication channel. Since the ledger is publicly available, we rely on public-key encryption to guarantee the secrecy of the decryption keys. In this case, each \Reader will publish a certified public key of a public-key encryption scheme at the beginning of the protocol execution. Thereby, each decryption key will be posted by each \Auth in an encrypted fashion to the ledger, so that the corresponding \Reader can later decrypt it with its own secret keys.
Our proof can be easily extended as follows. We may add a sequence of hybrids where we switch the encryptions of the decryption keys of $\formalMAABE$ made with the public-key encryption scheme to encryption of $0$. Then, we argue indistinguishability between each subsequent hybrid with a reduction to the CPA-security of the underlying public-key encryption scheme.%
\footnote{A similar technique is used by~\citet{BottaFVV21} to prove security of on-chain enabled general purpose multi-party computation protocols.}

Next, we describe our prototype serving as a proof of concept for MARTSIA, which we use to validate our approach in a blockchain-based process execution environment and against real-world applications.

%% file: sections/crypto-UC.tex
In the UC model of~\cite{Canetti01}, we distinguish between ideal and real execution.

\noindent{\emph{The Real Execution.}}
In the real world, the protocol $\pi$ is run in the presence of an adversary $\advA$ coordinated by a non-uniform environment $\advZ = \{\advZ_\secpar\}_{\secpar\in\NN}$.
At the outset, $\advZ$ chooses the inputs $(1^\secpar,\inp_i)$ for each player $\party_i$, and gives $\calI$, $\{\inp_i\}_{i\in\calI}$ and $z$ to $\advA$, where $\calI \subset [\nnum]$ represents the set of corrupted players and $z$ is some auxiliary input. For simplicity, we only consider static corruptions (i.e.,\ the environment decides who is corrupt at the beginning of the protocol).
The parties then start running $\pi$, with the honest players $\party_i$ behaving as prescribed in the protocol (using input $x_i$), and with malicious parties behaving arbitrarily (directed by $\advA$). The attacker may delay sending the messages of the corrupted parties in any given phase until after the honest parties send their messages in that phase. $\advZ$ can interact with $\advA$ throughout the course of the protocol execution.

Additionally, $\advZ$ receives the outputs of the honest parties, and must output a bit. We denote by $\REAL_{\pi,\advA,\advZ}(\secpar)$ the random variable corresponding to $\advZ$'s guess.

\noindent{\em The Ideal Execution.}
In the ideal world, a trusted third party evaluates the functionality $\Func$ on behalf of a set of dummy players $(\party_i)_{i \in [\nnum]}$. As in the real setting, $\advZ$ chooses the inputs $(1^\secpar,\inp_i)$ for each honest player $\party_i$, and gives $\calI$, $\{\inp_i\}_{i\in\calI}$ and $z$ to the ideal adversary $\advS$, corrupting the dummy parties $(\party_i)_{i \in \calI}$. Hence, honest parties send their input $x_i' = x_i$ to the trusted party, whereas the parties controlled by $\advS$ might send an arbitrary input $x_i''$. The trusted party behaves as prescribed by $\Func$.
During the simulation, $\advS$ can interact with $\advZ$ throughout the course of the protocol execution.
Additionally, $\advZ$ receives the outputs of the honest parties and must output a bit. We denote by $\IDEAL_{\Func,\advS,\advZ}(\secpar)$ the random variable corresponding to $\advZ$'s guess.

\begin{definition}[UC-Secure MPC]\label{def:mpc}
	Let $\pi$ be an $\nnum$-party protocol for computing a functionality $\Func$. We say that $\pi$ $\thr$-realizes $\Func$ in the presence of malicious adversaries if  for every PPT adversary $\advA$ there exists a PPT simulator $\advS$ such that, for every non-uniform PPT environment $\advZ$ corrupting at most $t$ parties, it holds that
	$
	\left\{\REAL_{\pi,\advA,\advZ}(\secpar)\right\}_{\secpar\in\NN} \cind \left\{\IDEAL_{\Func,\advS,\advZ}(\secpar)\right\}_{\secpar\in\NN}.
	$

\end{definition}

\ignore{
	\paragraph{Generalized UC.} An even stronger composability guarantee is that of \emph{generalized} UC~\cite{CanettiDPW07}. Roughly, in this setting, entities have black-box access to one or more global functionalities. A global functionality is an external functionality accessible by both the real and ideal world entities. The same instance of a global functionality can be accessed by multiple protocols (either simultaneously or sequentially depending on the model).
	
	\begin{definition}[GUC-Secure MPC]\label{def:mpc}
		Let $n\in\NN$.
		Let $\Func_f,\GFunc_g$ be ideal functionalities for $\func,\gunc:
		(\bin^*)^n \rightarrow (\bin^*)^n$, and let $\pi$ be an $n$-party protocol.
		We say that $\pi_f$ $t$-securely realizes in the $\GFunc_g$-GUC model if in the presence of malicious adversaries if  for every PPT adversary $\advA$ there exists a PPT simulator $\advS$ such that for every non-uniform PPT environment $\advZ$ corrupting at most $t$ parties the following holds:
		\[
		\left\{\REAL^g_{\pi,\advA,\advZ}(\secpar)\right\}_{\secpar\in\NN} \cind \left\{\IDEAL^g_{\func,\advS,\advZ}(\secpar)\right\}_{\secpar\in\NN}.
		\]
		where $g$ on top of the experiments $\REAL$ and $\IDEAL$ means that both experiments have black-box access to the function $g$.
	\end{definition}
	\paragraph*{The MPC Hybrid Model.}
	Let $\HYBRID^{\GFunc_1,\ldots,\GFunc_\tau}_{\pi,\advA,\advZ}(\secpar)$ denote the random variable 
	corresponding to the output of $\advZ$ in the $(\GFunc_1,\ldots,\GFunc_\tau)$-hybrid model.
	We say that a protocol $\pi$ for computing $\Func$ is secure in the $(\GFunc_1,\ldots,\GFunc_\tau)$-hybrid model if the ensembles $\HYBRID^{(\GFunc_1,\ldots,\GFunc_\tau)}_{\pi,\advA,\advZ}(\secpar)$ and $
	\IDEAL_{\Func,\Sim,\advZ}(\secpar)$ are computationally close.
}
\begin{sloppypar}
\begin{definition}[UC-Secure MPC in the hybrid model]\label{def:uc-hybrid}
	Let $n\in\NN$.
	Let $\Func$ and $\GFunc_1,\ldots,\GFunc_\tau$ be ideal functionalities,
	and let $\pi$ be an $n$-party protocol.
	We say that $\pi$ $\thr$-securely realizes $\Func$ in the $(\GFunc_1,\ldots,\GFunc_\tau)$-%
	hybrid model if for all PPT adversaries $\advA$ there exists a PPT simulator $\Sim$ 
	such that for all PPT non-uniform environments $\advZ$ corrupting at most $\thr$ 
	players, then
	$
	\left\{\IDEAL_{\Func,\Sim,\advZ}(\secpar)\right\}_{\secpar\in\NN} 
	\cind 
	\left\{\HYBRID^{\GFunc_1,\ldots,\GFunc_\tau}_{\pi,\advA,\advZ}(\secpar)\right\}_{\secpar\in\NN}.
	$
\end{definition}
\end{sloppypar}
\ignore{
	We further introduce the GUC variant of the MPC hybrid model.
	\begin{definition}[GUC-Secure MPC in the hybrid model]\label{def:uc-hybrid}
		Let $n\in\NN$.
		Let $\Func_\func$, $\Func_\gunc$, and $\GFunc_h$ be ideal functionalities for $\func,\gunc,h:
		(\bin^*)^n \rightarrow (\bin^*)^n$, and let $\pi$ be an $n$-party protocol.
		We say that $\pi$ $\thr$-securely realizes $\Func_\func$ in the $\Func_\gunc$-
		hybrid $\GFunc_h$-GUC model if for all PPT adversaries $\advA$ there exists a PPT simulator $\Sim$ 
		such that for all PPT non-uniform environments $\advZ$ corrupting at most $\thr$ 
		players, we have
		\begin{equation*}
			\left\{\IDEAL^h_{\func,\Sim,\advZ}(\secpar)\right\}_{\secpar\in\NN} 
			\cind 
			\left\{\HYBRID^{\gunc,h}_{\pi,\advA,\advZ}(\secpar)\right\}_{\secpar\in\NN}.
		\end{equation*}	
	\end{definition}
}

\noindent{\em Disjoint Parties Variant.}
In our setting, we are considering a slightly different corruption model taking into account multiple set of parties, in which the adversary have the power to corrupt at most a certain number of parties for each set.
We call a protocol with involved parties $\parties=(\parties_1,\ldots,\parties_m)$ where $|\parties_i|=n_i$ for $i\in[m]$ an $(n_1,\ldots,n_m)$-party protocol.

\begin{definition}[Disjoint Parties UC-Secure MPC in the hybrid model]\label{def:mult-parties-uc}
	Let $m\in\NN$ and $n_i\in\NN$ for $i\in[m]$. Let $\parties=(\parties_1,\ldots,\parties_m)$ where $|\parties_i|=n_i$ for $i\in[m]$.
	Let $\Func$ and $\GFunc_1,\ldots,\GFunc_\tau$, be ideal functionalities and let $\pi$ be an $(n_1,\ldots,n_m)$-party protocol.
	We say that $\pi$ $(\thr_1,\ldots,\thr_m)$-securely realizes $\Func$ in the $(\GFunc_1,\ldots,\GFunc_\tau)$-hybrid model if for all PPT adversaries $\advA$ there exists a PPT simulator $\Sim$ 
	such that for all PPT non-uniform environments $\advZ$ corrupting at most $\thr_i$ players in $\parties_i$ for $i\in[m]$, then
	$
	\left\{\IDEAL_{\Func,\Sim,\advZ}(\secpar)\right\}_{\secpar\in\NN} 
	\cind 
	\left\{\HYBRID^{\GFunc_1,\ldots,\GFunc_\tau}_{\pi,\advA,\advZ}(\secpar)\right\}_{\secpar\in\NN}.
	$
\end{definition}
\ignore{
	\begin{definition}[Disjoint Parties GUC-Secure MPC in the Hybrid Model]\label{def:mult-parties-guc}
		Let $m\in\NN$ and $n_i\in\NN$ for $i\in[m]$. Let $\parties=(\parties_1,\ldots,\parties_m)$ where $|\parties_i|=n_i$ for $i\in[m]$.
		Let $\Func_\func$, $\Func_\gunc$, and $\GFunc_h$ be ideal functionalities for $\func,\gunc,h:
		(\bin^*)^{\sum_{i\in[m]}n_i} \rightarrow (\bin^*)^{\sum_{i\in[m]}n_i}$, and let $\pi$ be an $(n_1,\ldots,n_m)$-party protocol.
		We say that $\pi$ $(\thr_1,\ldots,\thr_m)$-securely realizes $\Func_\func$ in the $\Func_\gunc$-
		hybrid $\GFunc_h$-GUC model if for all PPT adversaries $\advA$ there exists a PPT simulator $\Sim$ 
		such that for all PPT non-uniform environments $\advZ$ corrupting at most $\thr_i$ players in $\parties_i$ for $i\in[m]$, we have
		\begin{equation*}
			\left\{\IDEAL^h_{\func,\Sim,\advZ}(\secpar)\right\}_{\secpar\in\NN} 
			\cind 
			\left\{\HYBRID^{\gunc,h}_{\pi,\advA,\advZ}(\secpar)\right\}_{\secpar\in\NN}.
		\end{equation*}	
	\end{definition}
}

%% file: sections/crypto-ma-abe.tex
\begin{sloppypar}
We consider a ciphertext-policy MA-ABE scheme $\Pi =(\setup,\authsetup,\enc,\kgen,\dec)$ as 
defined by~\cite{Rouselakis} with an attribute universe $\attrs$, an authority universe $\auths$, a global identities universe $\idents$, a message space $\msgspace$, for any access structure $\pol$.
Since we are in the multi-authority setting, each attribute is associated with an authority. In particular, we consider mapping $\polpred: \attrs \rightarrow \auths$ from attributes to authorities.%
\footnote{For example, attributes can be structured as $\attr=(\attr',\auth)$, and $\polpred(\attr)$ outputs the second element of the pair.}
The algorithms are described as follows:

\begin{compactdesc}
	\item[$\setup(1^\secpar)$:] On input the security parameter $1^\secpar$, outputs the global parameters $\params$. We require that the attributes $\attrs$, authorities $\auths$, the global identifiers $\idents$, and the mapping $\polpred$ are included in $\params$.
	\item[$\authsetup(\params,\auth)$:] On input $\params$ and an authority index $\auth\in\auths$, outputs the authority's public key/secret key pair $\{\pk_\auth,\sk_\auth\}$.
	\item[$\kgen(\params,\id,\sk_\auth,\attr)$:] On input the global public parameters $\params$, the global identity $\id\in\idents$, the secret key $\sk_\auth$ of the authority $\auth\in\auths$ and an attribute $\attr\in\attrs$, outputs a decryption key $\dk_{\id,(\attr,\auth)}$ related to the attribute $\attr$ for the authority $\auth$.
	\item[$\enc(\params,\msg,\pol,\{\pk_\auth\}_{\auth\in\auths})$:] On input the global parameters $\params$, a message $\msg$, a policy $\pol$, public keys of all the authorities $\{\pk_\auth\}_{\auth\in\auths}$, outputs a ciphertext $\cipher$.
	\item[$\dec(\params,\cipher,\{\dk_{\id,\attr}\})$:] On input the global parameters $\params$, a ciphertext $\cipher$, a set of decryption keys $\{\dk_{\id,\attr}\}$ of a user $\id$ with respect to different attributes, outputs a message $\msg\in\msgspace$ or $\bot$.
\end{compactdesc}

\begin{definition}[Correctness]\label{def:maabe}
	A MA-ABE scheme (with ciphertext-policy) is correct if for any $\params$ generated by setup algorithm $\setup$, for any $\mathcal{L}\subseteq\auths$ and any pair of keys $\{\pk_\auth,\sk_\auth\}_{\auth\in\mathcal{L}}$ generated by the authority setup algorithm $\authsetup$, for any ciphertext $\cipher$ generated by the encryption algorithm $\enc$ using the relevant authorities’ public keys $\{\pk_\auth\}$ on any message $\msg$ and access structure $\pol$, and for any set of keys $\{\dk_{\id,\attr}\}$ generated by the key generation algorithm using the relevant authorities’ secret keys for one user $\id$ on any $\pol$-authorized set of attributes, we have that $\dec(\params,\cipher,\{\dk_{\id,\attr}\})=\msg$.
\end{definition}
\end{sloppypar}

\begin{definition}[CPA-security]\label{def:maabe-cpa}
	We say that an adversary statically breaks an MA-ABE scheme if it can guess the bit $b$ in the following security game with non-negligible advantage:
	
	\noindent{\textbf{Global Setup.}} The challenger runs the $\setup$ algorithms and sends the global parameters to the adversary.
	
	\noindent{\bf Attacker's queries.} The attacker answers as follows:
	\begin{asparaenum}
		\item \emph{Corruption query:} Chooses a set of corrupted authorities $\corr\subseteq\auths$ and the respective public keys $\{\pk_\auth\}_{\auth\in\corr}$, which can be maliciously crafted.
		\item \emph{Key Generation queries:} Sends polynomial number of queries of the type $\{(\id_i,\attrs_i)\}_{i\in[m]}$ where $\id_i\in\idents$, $\attrs_i\subseteq\attrs$, and $\polpred(\attrs_i)\cap\corr=\emptyset$.
		\item \emph{Challenge query:} Chooses two messages $\msg_0,\msg_1$ such that $|\msg_0|=|\msg_1|$ and a challenge access structure $\pol$. We require that for every key generation query submitted beforehand, $\attrs_i\cup\bigcup_{\auth\in\corr}\polpred^{-1}(\auth)$ is an unauthorized set for $\pol$, where $\bigcup_{\auth\in\corr}\polpred^{-1}(\auth)$ is the set of attributes belonging to the corrupted authorities.
	\end{asparaenum}
	\textbf{Challenger Replies.} The challenger calls $\authsetup(\params,\auth)$ for each $\auth\notin\corr$ and handles $\{\pk_{\auth}\}_{\auth\notin\corr}$ to the adversary. Then,
	flips $b\getsr\bin$ and replies as follows:
	\begin{asparaenum}
		\item \emph{Key generation queries response:} For each $(\id_i,\allowbreak\attrs_i)$ submitted by the adversary, compute and send $\kgen(\params,\allowbreak\id,\sk_{\polpred(\attr)})$ for each $\attr\in \attrs_i$.
		\item \emph{Challenge query response:} A ciphertext $\cipher^*\getsr\allowbreak\enc(\params,\allowbreak\msg_b,\pol,\{\pk_{\auth}\}_{\auth\in\auths})$. 
	\end{asparaenum}
	Finally, the adversary outputs a guess $b'$.
\end{definition}

%% file: sections/crypto-proofUC.tex
\begin{sloppypar}
\begin{proof}
We start by describing in \cref{fig:simulator} the simulator $\advS$ corrupting parties $\parties_3$, $\corr_1\subset\parties_1$, $\corr_2\subset\parties_2$, where $|\corr_1|\le n_2/2-1$ and $|\corr_2|\le n_3-1$, and
	interacting with an environment $\advZ$.  
	\begin{figure}[!h]
		\centering
		\begin{framed}
		\footnotesize
		\begin{description}[noitemsep]
			\item[Setup:] Initialize a dictionary $\dict$. Acting as $\Func_\setup$, compute $\params\getsr\setup(1^\secpar)$, and forward $\params$ to $\advZ$. Then, acting as $\FuncSC$, upon receiving $(\SetAttr,\attrs,\certattrs)$ from $\advZ$, forward $(\SetAttr,\attrs,\certattrs)$ to $\advZ$. Then, Wait to receive $\{(\params',\pk_{\bar\auth})\}_{\bar{\auth}\in\corr_1}$ from $\advZ$.  Finally, for each $\auth\in\parties_1\setminus\corr_1$, compute $\sk_\auth\getsr\authsetup(\params,\auth)$.
			\item[Key Generation:]
			For each $(\reader,\attr)\in\certattrs$ where $\reader\in\corr_1$ and $\polpred(\attr)\notin \corr_1$ compute $\dk_{\reader,\attr}\getsr\kgen(\params,\reader,\sk_{\polpred(\attr)},\attr)$ for each $\attr\in \attrs$, and send $\{\dk_{\reader,\attr}\}_{\attr\in \bar{\attrs}}$ to the adversary.
			\item[Encryption:] 
			Upon receiving $(\messagestored,\pol)$ from $\FuncTR$:
			
			(1) If there exists least a reader $\reader\in\corr_1$ for which the attributes in a subset
			$\{(\reader,\attr)\}$ where $(\reader,\attr)\in\certattrs$  compose an authorized set $\bar{\attrs}$ of $\pol$, query $\FuncTR$ with $(\readmessage,\bar{\attrs})$, obtaining back a set of messages $\messages$. Then, for each message $\msg\in\messages$, compute $\cipher\getsr\enc(\params,\msg,\pol,\{\pk_\auth\}_{\auth\in\auths})$.  
			(2)   Else, compute $\cipher\getsr\formalMAABE.\enc(\params,0,\pol,\{\pk_\auth\}_{\auth\in\auths})$.
			(3) Set $\dict[\pol]\gets\dict[\pol]\cup\{\cipher\}$.

			\item[Decryption:] Acting as $\FuncSC$, upon receiving $(\RetrieveCtx,\tilde{\attrs})$ from $\advA$ on behalf of $\reader$ such that $(R,\alpha)\in\mathcal{V}$ for each $\alpha\in\tilde{\attrs}$, set $\ctxs=\emptyset$ and for each policy $\pol\in\dict$ such that $\tilde{\attrs}$ is an authorized set for $\pol$,
			set $\ctxs\gets\ctxs\cup\dict[\pol]$. Finally,  output $(\RetrieveCtx,\ctxs)$ to $\advZ$.
			\item[Dictionary Retrieval:]  When receiving $(\RetrieveDict)$ from $\advZ$, output $(\RetrieveDict,\dict)$ to $\advZ$.
		\end{description}
	\end{framed}
	\caption{Simulator $\advS$ of $\IDEAL_{\FuncTR,\advS,\advZ}(\secpar)$}
	\label{fig:simulator}
\end{figure}
Let us assume that $\advZ$ distinguishes $\HYBRID^{\Func_\setup,\FuncSC}_{\pitr,\advA,\advZ}(\secpar)$ and 
$\IDEAL_{\FuncTR,\advS,\advZ}(\secpar)$ with non-negligible probability, then we can construct an adversary 
$\advabe$ breaking the CPA-security of $\formalMAABE$ (\cref{def:maabe-cpa}).  Let us assume that $\dataowner$ encrypts only one message with a policy for which the adversary is unauthorized. $\advabe$ behaves as follows:
\begin{compactdesc}
	\item[Setup:] Initialize a dictionary $\dict$, compute $\params\getsr\setup(1^\secpar)$ and forward $\params$ to $\advZ$. Then, collect all the corrupted authorities' public keys $\{\pk_{\bar{\auth}}\}_{\bar{\auth}\in\corr_1}$ from $\advZ$ and forward them to the challenger.
	\item[Key generation:] Collect all the key generation requests $\{(\reader_i,\attrs_i)\}_{i\in[\kappa]}$ from $\advZ$ such that $\polpred(\attrs_i)\cap\corr_1=\emptyset$, and forward them to the challenger, receiving back $\{\dk_{\reader_i,\attr}\}_{i\in[\tau],\attr\in\attrs_i}$. Forward $\{\dk_{\reader_i,\attr}\}_{i\in[\kappa],\attr\in\attrs_i}$ to $\advZ$.
	\item[Encryption:]  When receiving $(\messagestored,\pol)$ from $\FuncTR$,
	(1) If there exists at least a reader $\reader\in\corr_1$ for which the attributes in a subset
	$\{(\reader,\attr)\}$ where $(\reader,\attr)\in\certattrs$  compose an authorized set $\bar{\attrs}$ of $\pol$, query $\FuncTR$ with $(\readmessage,\bar{\attrs})$, obtaining back a set of messages $\messages$. Then, for each message $\msg\in\messages$, compute $\cipher\getsr\enc(\params,\msg,\pol,\{\pk_\auth\}_{\auth\in\auths})$.  
	(2)    Else, flip a random bit $b\gets\bin$. 
	If $b=0$, set $\cipher\getsr\formalMAABE.\enc(\params,\msg,\allowbreak\pol,\{\pk_\auth\}_{\auth\in\auths})$, where $\msg$ is the real message. Else,  set  $\cipher\getsr\formalMAABE.\enc(\params,0,\pol,\allowbreak\{\pk_\auth\}_{\auth\in\auths})$.
	(3) Set $\dict[\pol]\gets\dict[\pol]\cup\{\cipher\}$.
	\item[Decryption:] When receiving $(\RetrieveCtx,\tilde{\attrs})$ from $\advA$ on behalf of $\reader$ such that $(R,\alpha)\in\mathcal{V}$ for each $\alpha\in\tilde{\attrs}$, set $\ctxs=\emptyset$ and for each policy $\pol\in\dict$ such that $\tilde{\attrs}$ in an authorized set for $\pol$,
	set $\ctxs\gets\ctxs\cup\dict[\pol]$. Finally,  output $(\RetrieveCtx,\ctxs)$ to $\advA$.
	\item [Dictionary retrieval:]When receiving $(\RetrieveDict)$ from $\advZ$, output $(\RetrieveDict,\dict)$ to $\advZ$.
\end{compactdesc}
If $b=0$, $\advabe$ perfectly simulates $\HYBRID^{\FuncSC,\Func_\setup}_{\pitr,\advA,\advZ}(\secpar)$. Whereas if $b=1$ $\advabe$ perfectly simulates $\IDEAL_{\FuncTR,\advS,\advZ}(\secpar)$. Note that when the adversary is authorized by $\pol$ (hence he previously obtained decryption keys related to attributes decrypting the ciphertexts given by $\dataowner$ for $\pol$),  the simulator can retrieve all the messages that $\dataowner$ sent to $\FuncTR$  by simply querying $\FuncTR$ with $(\readmessage,\pol)$ on behalf of the authorized corrupted reader (step 3 of the encryption phase). Hence, the distribution of such ciphertexts  in the real execution and the ideal execution will be identical.
On the contrary, when the adversary is unauthorized by $\pol$ (step 2 of the encryption phase), the ideal and the real execution are computationally indistinguishable due to the underlying CPA-security of $\formalMAABE$. This holds since the condition that $\attrs_i\cup\bigcup_{\auth\in\corr}\polpred^{-1}(\auth)$ is an unauthorized set of $\pol$ is always satisfied because we require that $\dataowner$ specifies policies where each attribute is linked with the majority of authorities, and hence the adversary, corrupting a minority of the authorities will never be able able to authorize himself by generating corrupted decryption keys. Now, the adversary $\advZ$ obtains a ciphertext (fetched with the $(\RetrieveDict)$ query) that is either an encryption of the real message given by $\dataowner$ in the real world or an encryption of $0$. If $\advZ$ can distinguish between the real and the ideal world, he can guess which of the two messages was encrypted with non-negligible probability, thus breaking the CPA-security of $\formalMAABE$.
We can conclude that  $\HYBRID^{\FuncSC,\Func_\setup}_{\pitr,\advA,\advZ}(\secpar)\approx_c \IDEAL_{\FuncTR,\advS,\advZ}(\secpar)$.

\noindent
The proof can be trivially extended to the case in which $\dataowner$ encrypts multiple messages for different access policies where the adversary is not authorized. To do so, we can rely on a hybrid argument where, in each subsequent hybrid, we change the encryption of the $i$-th message related to a policy for which the adversary is not authorized, with the encryption of $0$. Indistinguishability between subsequent hybrids can be proven by repeating the above reduction. 
%
\end{proof}
\end{sloppypar}

%% file: sections/impl-eval.tex
We have hitherto discussed the functionalities and security guarantees of MARTSIA by design, both by the explanation of the components and their interactions that underpin those properties, and through a formal analysis. 
In this section, we illustrate our proof-of-concepts implementation of MARTSIA over multiple platforms (\cref{sec:swimpl}, in compliance with \Req{9}). Thereafter, we demonstrate the practical integrability of MARTSIA with an existing state-of-the-art blockchain-based workflow engine (\cref{sec:integration:wfengine}, demonstrating compliance with requirements from \Req{1} to \Req{6}), assess the prototype's performance (\cref{sec:perfomance:analysis}, with insights into \Req{7}~and~\Req{8}), and its adoption in real-world key application domains (\cref{sec:realworld:dapps}).

\subsection{Software implementation} \label{sec:swimpl}
To this end, we implemented the {\SC}s described in \cref{sec:approach:workflow} in Solidity v.~0.8.20 for the \EVM and in PyTEAL v.~0.20.0%
\footnote{See \href{https://docs.soliditylang.org/}{\nolinkurl{docs.soliditylang.org}} and \href{https://pyteal.readthedocs.io/}{\nolinkurl{pyteal.readthedocs.io}}.}
for the \AVM (see \cref{sec:background}). 
%
%
We deployed our software objects on the Sepolia (Ethereum), Mumbai (Polygon), and Fuji (Avalanche) \EVM testnets,%
\footnote{\label{foot:evm:testnets}%
	Sepolia: \href{https://sepolia.etherscan.io/}{\nolinkurl{sepolia.etherscan.io}}; Mumbai: \href{https://www.oklink.com/mumbai}{\nolinkurl{oklink.com/mumbai}}; Fuji: \href{https://testnet.snowtrace.io/}{\nolinkurl{testnet.snowtrace.io}}. Accessed: October 30, 2024.%
}
and on the Algorand \AVM testnet.
\footnote{Algorand testnet: \href{https://testnet.explorer.perawallet.app/}{\nolinkurl{testnet.explorer.perawallet.app/}}. Accessed: October 30, 2024.}
Notice that, though being both public permissionless blockchains supporting the programmability of smart contracts, Algorand and Ethereum apply different paradigms in their protocols. Among the various distinctive characteristics, we recall, e.g., that Algorand natively supports multi-sig accounts, whereas Ethereum does not. However, both can underpin MARTSIA in light of its cross-platform nature, as per \Req{9}.
To implement the \DStore, we have created an \IPFS local node.%
\footref{foot:ipfs}
Furthermore, we have realized the off-chain components backing the system (e.g., the client-server communication channels) in Python.
The code of our prototypes, alongside the detailed results of our experiments, can be found at
\href{https://github.com/apwbs/MARTSIA}{\nolinkurl{github.com/apwbs/MARTSIA}}. 

\begin{figure*}[tbp]
\centering
	\includegraphics[width=\textwidth]{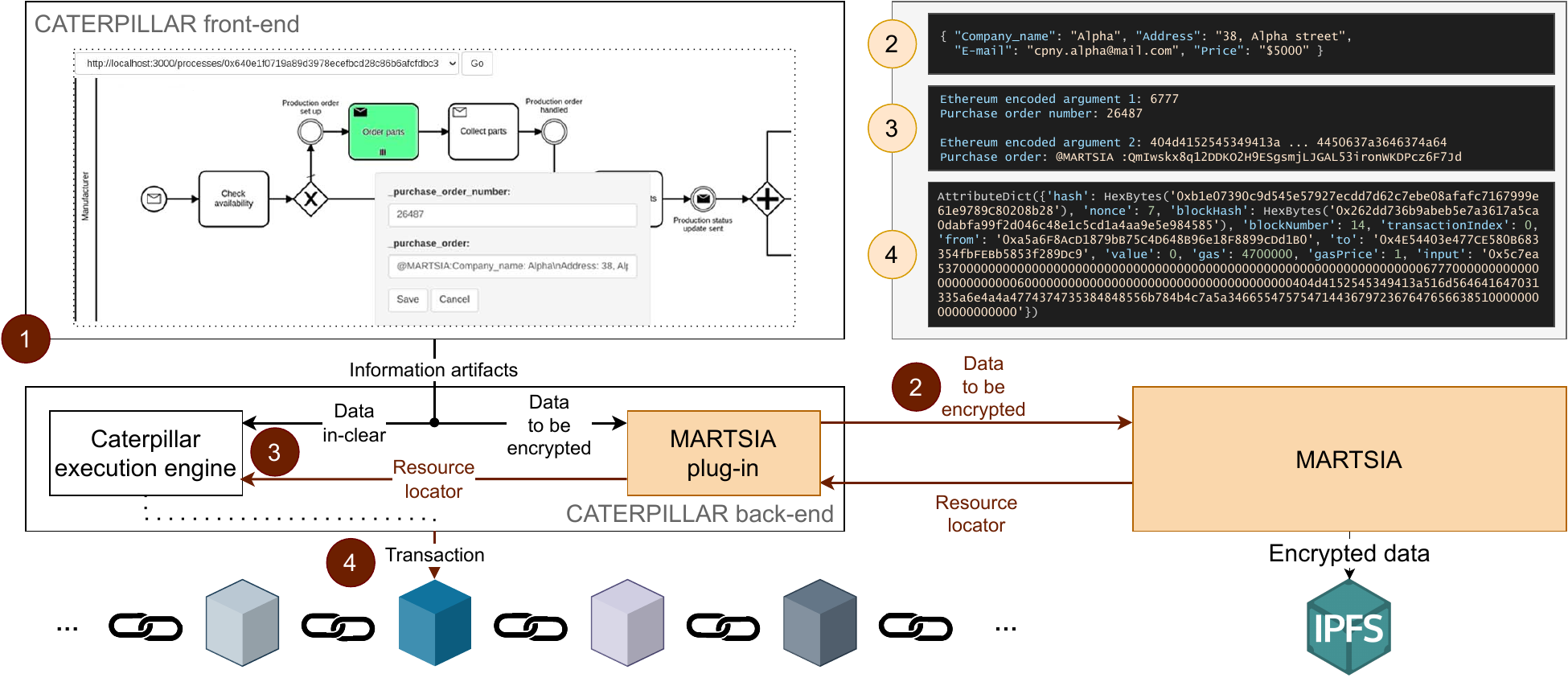}
	\caption{A run of our integration of MARTSIA with Caterpillar~\citep{Lopez-Pintado.etal/SPE2019:Caterpillar}}
	\label{fig:martsia-x-caterpillar}
\end{figure*}

\subsection{Integration with a process execution engine}
\label{sec:integration:wfengine}
First, to demonstrate the adoption of MARTSIA as a secure data-flow management layer for process execution (thus meeting requirement \Req{6}), we created an integration module connecting our tool with a state-of-the-art blockchain-based process execution engine tool, i.e., Caterpillar~\citep{Lopez-Pintado.etal/SPE2019:Caterpillar}. 
For our experiments, we used Caterpillar v.1.0.%
\footnote{\href{https://github.com/orlenyslp/Caterpillar}{\nolinkurl{github.com/orlenyslp/Caterpillar}}. Accessed: October 30, 2024.}
The code to replicate and extend our test is publicly available in our repository.%
\footnote{\href{https://github.com/apwbs/MARTSIA-Ethereum/tree/main/caterpillar-interaction}{\nolinkurl{github.com/apwbs/MARTSIA-Ethereum/tree/main/caterpillar-interaction}}}
As shown in \cref{fig:martsia-x-caterpillar}, we insert a plug-in in the architecture of Caterpillar to use MARTSIA as an intermediate data layer manager, replacing the built-in data store for information securing.
We illustrate our experiment with a simplified fragment of the ru¨nning example (see \cref{sec:example}) focusing on the purchase order sent from the manufacturer to the international supplier. The exchanged data artifact consists of a purchase order number, to be publicly stored on chain, and the confidential purchase order entries listed in the top row of \cref{tab:messageEncoding:before}. The user passes both the entries as input through the Caterpillar panel (see mark~{\BCircOne} in \cref{fig:martsia-x-caterpillar}), specifying the data they want to be secured by MARTSIA with a special prefix (``\texttt{@MARTSIA:}'').
Our integration module captures the input and encrypts the indicated entry as explained in \cref{sec:approach} so that only the supplier can 
interpret those pieces of information as per \Req{1} (\BCircTwo). After the encryption, 
MARTSIA invokes the Caterpillar Smart Contract 
passing as the second argument 
an \IPFS link in place of the original data (\BCircThree). The resource locator for the stored information is thus saved on the ledger by the process execution engine for future audits (\BCircFour), yet not publicly readable (\Req{2}). Thereupon, the recipient of the confidential information (or the auditor, later on) can retrieve and decode the information with their secret key (\Req{4}) provided by the authority network (\Req{5}). 

Aside from empirically showing the suitability of MARTSIA as a secure data-flow manager in an ensemble with a process execution engine, we remark that the cost overhead in terms of transaction fees required by MARTSIA is negligible with respect to the main process execution management. For example, the method running the activity \texttt{Order parts} of the BPMN in~\cref{fig:example}
on Caterpillar incurs \num{114494} gas units for execution with our inputs.
The on-chain component of MARTSIA, detached from Caterpillar, requires \num{89772} gas units to store the \IPFS link. As we use the second input field of Caterpillar's smart contract to save that resource locator, the separate gas consumption for MARTSIA is unnecessary and can be directly included in the overall process execution costs. Notice that the same activity execution saving the purchase order as plaintext (thus renouncing to the fine-grained confidentiality guarantees of MARTSIA) would 
have entailed a larger cost as the textual content is a longer string than an \IPFS link: \num{116798} gas units.
Auditability on process execution \emph{and} secure message exchange are thus guaranteed with low overhead, as per \Req{3}.

\subsection{Cost analysis}
\label{sec:cost:analysis}
To gauge the gas expenditure and execution time of our system's on-chain components, we called the methods of the deployed {\SC} daily on Sepolia, Fuji and Mumbai\footref{foot:evm:testnets} for \num{14} days, from 16 to 29 May 2023.
%
\begin{table}[tb]
	\caption[Execution cost and timing of the steps that require an interaction with the blockchain]{Execution cost and timing of the steps that require an interaction with EVM blockchain platforms: Ethereum (ETH, via the Sepolia testnet), Avalanche (AVAX, via the Fuji testnet), and Polygon (MATIC, via the Mumbai testnet).}%
	\label{tab:exec:cost}%
	\resizebox{\textwidth}{!}{
		\input{tables/execution-cost-table}
	}
\end{table}
The data we used to run the tests stems from our running example (see \cref{sec:example}).
\Cref{tab:exec:cost} illustrates the results.
We perform the measurement in six phases: 
\begin{iiilist}
	\item the deployment of the \SC;
	\item the initialization of the {\Auths} (steps~\textbf{2.\ref{step:authinit:metadata}}~to~\textbf{2.\ref{step:authinit:pkpk}});
	\item the {\Reader} certification (step~\textbf{3.\ref{step:key:writeattrlinkonchain}});
	\item the key request of a \Reader to an \Auth (step~\textbf{3.\ref{step:key:request}}), and
	\item the return of the key in response (step~\textbf{3.\ref{step:key:return}}); 
	\item the storage of a message by a \DOwner to save a message (step~\textbf{4.\ref{step:dataex:policyenconchain}}).
\end{iiilist}
For all the above phases, the table shows the average gas consumed for execution in gas units (abbreviated as ``\GasU'' in the table, ranging from \num{21848} for step~\textbf{3.\ref{step:key:request}} to \num{1692955} for the \SC deployment) and the cost converted in Gwei (i.e., $10^{-9}$ Ether).

Notice that we have two columns associated to the deployment phase \textit{(i)}, labeled with ``SC deployment (full)'' and ``SC deployment (w/o \Req{7} \& \Req{8})'', respectively. The latter indicates a leaner implementation wherein the deployment and access-grant updates are complete upon a single invocation, as opposed to the former demanding that the majority of {\AttCert}s confirm the operations. 
On an \EVM, the former requires \num{3870505} units of Gas overall, i.e., \SI{128.6}{\%} more than the latter.
As multi-sig wallets are implemented natively in Algorand (see \cref{sec:background}), the deployment costs are equal on an \AVM. Therefore, meeting \Req{7} and \Req{8} comes at no additional costs. 
Since we use local variables to manage the roles (\eg authorities, users), the minimum balance needed to deploy these new \SC does not change (\num{0.2785} Algos). 
However, agents who want to interact with the system must opt-in to the full \SC. This increases their minimum balance up to a maximum of \num{0.1785} Algos. 
We recall that these minimum balance increments do not constitute additional payments per se: these funds are locked and can be unlocked upon the opt-out (see \cref{sec:background}).



To have an estimation of the expenditures on the mainnets, we operated as follows.
We collected the daily minimum gas price (i.e., the minimum price of the gas of a transaction included in the blocks) on the Ethereum, Polygon, and Avalanche mainnets for six months, from the 1\textsuperscript{st} of January to the 1\textsuperscript{st} of June 2023. 
This period's minimum gas price in Gwei oscillated between \num{9.26} and \num{90.01} for Ethereum, between \num{0.00078} and \num{0.14} for Polygon, and between \num{0.19} and \num{0.28} for Avalanche.
Considering the most and least expensive mainnets (i.e., Ethereum and Polygon, respectively), the most complex operation (i.e., the deployment of the full {\SC}s) would have required at least between \num{35837772,15} (minimum) and \num{348390971.01} (maximum) Gwei on Ethereum, and between \num{3011.25} and \num{554666,59} Gwei on Polygon. Conversely, the least complex operation (i.e., the key return at Step 3.~\ref{step:key:return}) would have demanded at least between \num{204369.02} and \num{1986739.58} Gwei on the former, and between \num{17.17} and \num{3163.05} Gwei on the latter.
We remark that the cost analysis in Algorand is straightforward as 
each transaction costs \num{0.001}~Algos 
regardless of the types (\eg a {\SC} deployment or a method invocation).

\subsection{Performance analysis}
\label{sec:perfomance:analysis}
Along with the costs, we 
measured the time needed to run our system in a stepwise fashion. Each step involves sending a transaction to the blockchain. Therefore, we separate the execution time between the off-chain data elaboration and the latency induced by the blockchain infrastructure. The results are reported in the bottom line of \cref{tab:exec:cost}. The average time required to store a transaction in a block ranges from approximately \SI{4.3}{\sec} (Fuji) to \SI{9.3}{\sec} (Sepolia) for \EVM-based platforms, and \SI{4.4}{\sec} for Algorand. 
The off-chain passages require a lower time, from about \SI{0.038}{\sec} for the {\Reader} certification ({Step~\textbf{1.\ref{step:key:writeattrlinkonchain}}}) to the circa \SI{2.6}{\sec} needed for the cooperative work carried out by the {\Auths} during the initilization phase (steps~\textbf{0.\ref{step:authinit:metadata}}~to~\textbf{0.\ref{step:authinit:pkpk}}).
More in-depth comparative analyses and a stress test of the architecture 
pave the path for future endeavors, as we discuss in \cref{sec:conclusion} after a summary of the state of the art.

\begin{figure*}[tbp]
	\includegraphics[width=\textwidth]{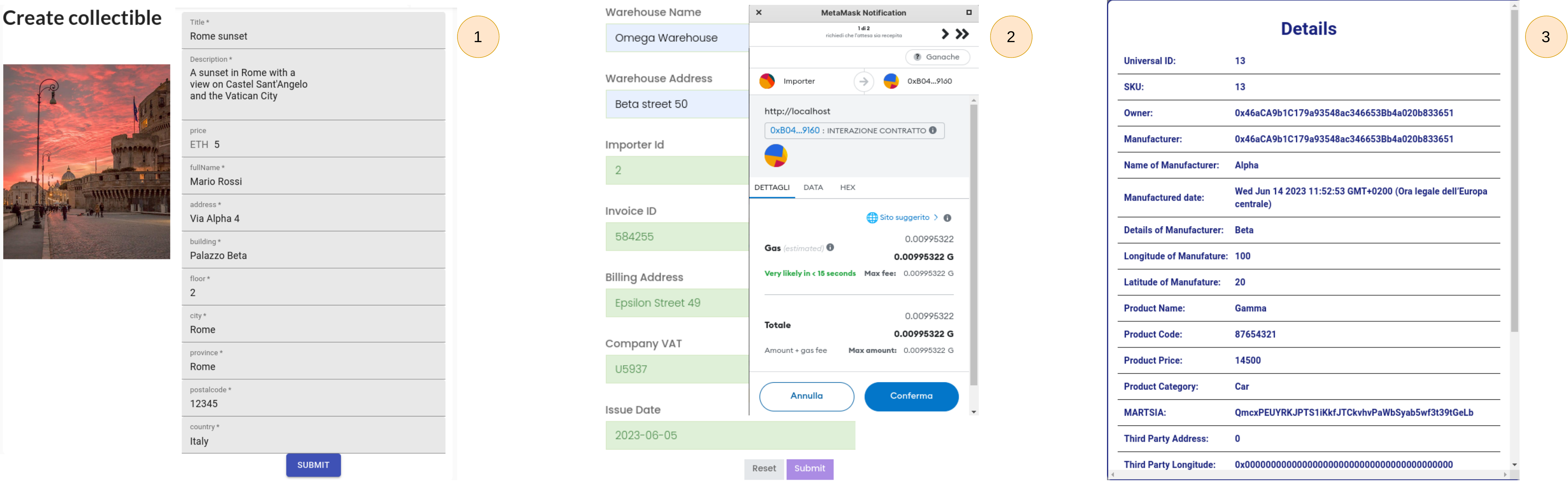}
	\caption{Interaction between MARTSIA and three DApps: {\CircOne} NFT Marketplace; {\CircTwo} Coffee supply chain; {\CircThree} Retail.}
	\label{fig:dapp-interaction}
\end{figure*}
\subsection{Integration with real-world applications}
\label{sec:realworld:dapps}
To showcase the use of MARTSIA in key domains for blochchain technologies, we have integrated our prototype with three publicly available, renown \DApps, each
representing
a key use case:
{\CircOne} \NFT trading~\citep{schaar2022non,DBLP:conf/icdcs/MorgiaMMN23}, {\CircTwo} supply-chain management~\citep{dutta2020blockchain}, and {\CircThree} retail~\citep{tijan2019blockchain}.
In the following, we enter the detail of the three case studies we selected, highlighting the gap left by a lack of a specific component managing the access control over exchanged data. \Cref{fig:dapp-interaction} shows screenshots of the platforms enhanced with plug-ins to interface the existing \DApps with MARTSIA.%

\noindent{\CircOne}
	The first \DApp is an \NFT marketplace that allows for the creation, sale and purchase of pieces of arts.%
	\footnote{\href{https://github.com/BravoNatalie/NFT-Marketplace}{\nolinkurl{github.com/BravoNatalie/NFT-Marketplace}}. Accessed: October 30, 2024.}
	It enables artists to upload digital versions of their artwork, together with the title, description, and the NFT price in ETH. Additional information that pertains to the \NFTs and the paired asset but should not be disclosed to the general public are 
	the artist's name, and the full address of the warehouse where the original artwork is kept, which should be visible only to specific actors like 
	the artist and 
	the art gallery curators. 

\noindent{\CircTwo}
	The second use case focuses on a coffee supply chain, from the cultivation of raw coffee beans to the transformation into processed coffee. 
	This implementation ensures the establishment of an immutable and verifiable data source.%
	\footnote{\href{https://github.com/rwaltzsoftware-org/coffee-supplychain-ethereum}{\nolinkurl{github.com/rwaltzsoftware-org/coffee-supplychain-ethereum}}. Accessed: October 30, 2024.}
	Several parties are engaged in the supply chain process. 
	An admin creates batches. A farm inspector provides general coffee information. A harvester processes the coffee and updates the records on temperature and humidity. Exporter and importers handle the information about coffee logistics. Finally, processors roast beans for packaging and sale.
	During these steps, all recorded details are fundamental
	to verify that all the stages that compose the supply chain are performed correctly giving immutable and verified data. However, parts of the data must remain secret to the majority of the actors involved. For example, the importer's 
	invoice and its identifier, the billing address, the company VAT and the date should remain visible only to the importer, the exporter and a third entity 
	(\eg a government organization) 
	that verifies law compliance. 

\noindent{\CircThree}
	The last showcase is a \DApp wherein customers can purchase products by a retail merchant. 
	The blockchain ensures a transparent and secure transfer of the traded goods.%
	\footnote{\href{https://github.com/rishav4101/eth-supplychain-dapp/tree/main}{\nolinkurl{github.com/rishav4101/eth-supplychain-dapp/tree/main}}. Accessed: October 30, 2024.}
	In this \DApp, the workflow is as follows. First, a manufacturer adds a product to the platform and a third party buys it. 
	Once the manufacturer has shipped the product to a third party, a customer can purchase it. Once the transaction is completed, the third party ships the product to the delivery hub. At this stage, the product is verified and shipped 
	to the customer who has bought it. Every step 
	is publicly available on a blockchain. 
	However, other relevant details should be kept though not available to the general public. When the manufacturer adds a new product to the platform, e.g., the data about the CSDD 
	should only be readable by the manufacturer, 
	the third party, and the delivery hub. 
	
	

For all of the three cases above, we provided a few key confidential pieces of information (e.g., the name of the artwork's author in the first case, the importer's invoice in the second case, and the CSDD in the third case) that should be appended to the exchanged data artifacts. Still, the original \DApps cannot support access control over the exchanged data. Therefore, they renounce to the recording of those that should be confidential at once, rather than exposing them in-clear. 
This, however, reduces consistency and auditability as confidential information is detached from the blockchain-based notarization and scattered among different, possibly falsifiable means. 
To solve this issue, we have extended the input interfaces to accept additional reserved information -- see, e.g., \cref{fig:dapp-interaction}~({\CircOne}) from the fourth text field ``fullName'' down to ``country''. Thereupon, we adopt a similar scheme to the Caterpillar case above illustrated in~\cref{fig:martsia-x-caterpillar}: We endow the \DApps with a tailored plug-in that intercepts the user-entered information -- see the transaction confirmation step in~\cref{fig:dapp-interaction}~({\CircTwo}). What requires access control is transferred to MARTSIA, which finally returns to the \DApp the \Rloc (IPFS link) of the ciphered data -- see the field labeled with ``MARTSIA:'' in~\cref{fig:dapp-interaction}~({\CircThree}). Public data continues the natural flow in the \DApp, thus guaranteeing retro-compatibility with the original applications.

%% file: tables/execution-cost-table.tex
\begin{tabular}{l
  S[table-format = 6.3,round-precision = 3,round-mode=places]
  S[table-format = 6.3,round-precision = 3,round-mode=places]
  S[table-format = 7.3,round-precision = 3,round-mode=places]
  S[table-format = 5.3,round-precision = 3,round-mode=places]
  S[table-format = 5.3,round-precision = 3,round-mode=places]
  S[table-format = 5.3,round-precision = 3,round-mode=places]
  S[table-format = 6.3,round-precision = 3,round-mode=places]
  S[table-format = 4.3,round-precision = 3,round-mode=places]
}
    \toprule
      {~} &
      \multicolumn{7}{c}{\textbf{Execution cost} $\left[\mathrm{Gwei}=\mathrm{ETH} \times10^{-9}\right]$} &
      \\
      \cmidrule{2-8}
      {~} &
      \multicolumn{1}{c}{{SC depl.\ (full)}} & 
      \multicolumn{1}{r}{{SC depl.\ (w/o \Req{7}\&\Req{8})}} & 
      \multicolumn{1}{c}{{Steps~\textbf{2.\ref{step:authinit:metadata}}~to~\textbf{2.\ref{step:authinit:pkpk}}}} & 
      \multicolumn{1}{c}{{Step~\textbf{3.\ref{step:key:writeattrlinkonchain}}}} & 
      \multicolumn{1}{c}{{Step~\textbf{3.\ref{step:key:request}}}} &
      \multicolumn{1}{c}{{Step~\textbf{3.\ref{step:key:return}}}} &      
      \multicolumn{1}{c}{{Step~\textbf{4.\ref{step:dataex:policyenconchain}}}} &
      \multicolumn{1}{c}{\textbf{Avg.\ latency}}
      \\ 
      \multicolumn{1}{l}{\textbf{Platform}} &
      \multicolumn{1}{c}{(\num{3870505} \GasU)} &
      \multicolumn{1}{c}{(\num{1692955} \GasU)} &
      \multicolumn{1}{c}{(\num{476547} \GasU)} &
      \multicolumn{1}{c}{(\num{67533} \GasU)} &
      \multicolumn{1}{c}{(\num{21848} \GasU)} &
      \multicolumn{1}{c}{(\num{22072} \GasU)} &
      \multicolumn{1}{c}{(\num{89772} \GasU)} &
      \multicolumn{1}{c}{[\si{\milli\second}]}\\
      \cmidrule(r){1-1}
      \cmidrule{2-8}
      \cmidrule(l){9-9}
      \textbf{Sepolia} &
       5805142.727004e-09 &
       2539432.51354364e-09 &
       714820.503880085e-09 &
       101299.500540264e-09 &
       32748e-09 & 
       33084e-09 & 
       134658.000718176e-09 &
       9288.573646942760 \\
      \textbf{Fuji} &
       778380.190506e-09 &
       340498.77068816e-09 &
       95873.484581149e-09 &
       13586.538284745e-09 &
       4392.91e-09 & 
       4437.98e-09 & 
       18060.662362105e-09 &
       4278.099486033120 \\
      \textbf{Mumbai} &
       2933.310618e-09 &
       1283.162694703e-09 &
       354.691115093e-09 &
       50.311010581e-09 &
       16.05e-09 & 
       16.21e-09 & 
       66.011515441e-09 &
       4944.807272752120 \\
       \midrule
       &
       \multicolumn{7}{c}{\textbf{Off-chain execution time}~{[\si{\milli\second}]}}
       & \\
       \cmidrule{2-8}
       &
       0.000 &
       0.000 &
       2582.47053623199 &
       38.27977180481 &
       134.573 &
       714.011 &
       158.4467888 &
       \\
    \bottomrule
\end{tabular}

%% file: sections/sota.tex
In recent years, numerous approaches have been proposed to automate collaborative processes using blockchain
technology~\citep{DiCiccio.etal/InfSpektrum2019:BlockchainSupportforCollaborativeBusinessProcesses} beyond the aforementioned Caterpillar~\citep{Lopez-Pintado.etal/SPE2019:Caterpillar}.
Previous studies in the area have shown the effectiveness of blockchain-based solutions to add a layer of trust among actors in multi-party collaborations~\citep{Weber.etal/BPM2016:UntrustedBusinessProcessMonitoringandExecutionUsingBlockchain} even in adversarial settings~\citep{Madsen.etal/FAB2018:CollaborationamongAdversaries:DistributedWorkflowExecutiononaBlockchain}, improve verifiability of workflows with model-driven approaches~\citep{Lopez-Pintado.etal/SPE2019:Caterpillar,Tran.etal/BPMDemos2018:Lorikeet}, allow for monitoring~\citep{DiCiccio.etal/SoSyM2022:BlockchainForProcessMonitoring}, mining~\citep{Klinkmueller.etal/BCForum2019:ExtractingProcessMiningDatafromBlockchainApplications}, security~\citep{Koepke.etal/FGCS2023:DesigningSecureBusiness}, and auditing~\citep{Corradini.etal/ACMTMIS2022:EngineeringChoreographyBlockchain,Corradini.etal/BCRA2021:ModelDrivenEngineering}.
Interestingly, a more recent release of Caterpillar~\citep{Lopez-Pintado.etal/IS2022:ControlledFlexibilityBlockchainCollaborativeProcesses} enables the dynamic allocation of actors based on a language for policy bindings. 
MARTSIA has the capability to adjust roles dynamically as well, as access keys are created based on actors' attributes verified at runtime. 
These studies enhance the integration of blockchain technology with process management, unlocking security and traceability benefits. However, they primarily focus on the control-flow perspective and lack mechanisms for secure access control to the stored data on public platforms. Our work focuses right on this key aspect of collaborative business and, as demonstrated in \cref{sec:imptes}, can complement existing blockchain-based process execution engines.

Another area of research related to our investigation is the preservation of data privacy and integrity in a blockchain system.
%
Several papers in the literature explore the use of encryption for this purpose. Next, we provide an overview of techniques. 
%
Hawk~\citep{Hawk} is a decentralized system that utilizes user-defined private Smart Contracts to automatically implement cryptographic methods.
\citet{henry2022random} employ smart contracts that handle payment tokens. Banks operate as trustworthy intermediaries to preserve privacy. MARTSIA pursues confidentiality of exchanged information too, although it does not resort to central authorities (the banks) to this end. Also, our approach resorts to smart contracts too for notarization purposes but does not require the encoding of custom or domain-specific ones. 

\citet{RZKPB} introduce RZKPB, a privacy protection system for shared economy 
based on blockchain technology. Similarly to MARTSIA, this approach does not involve third parties and resorts to external data stores.
Differently from their approach, we link data on chain with the data stores so as to permanently store the resource locators.
\todo{ACM TOPS added}
\citet{Flexichain} propose Flexichain, an off-chain flexible payment channel network protocol. Flexichain allows users to deposit coins per user rather than per channel. In this solution the users deposit a certain amount of coins to the blockchain and then all the computations are executed off-chain. In our solution, we adopt a similar approach since we perform off-chain encryption and then publish on-chain only the result of the computation.
\citet{BenhamoudaCanAP} introduce a solution that enables a public blockchain to serve as a storage place for confidential data.
As in our approach, they utilize shared secrets among components. However, their approach discloses the secret when determined conditions are fulfilled, whereas MARTSIA does not reveal secret data on the blockchain at any stage.
\todo{Citation of ACM TOPS suggested editor: added}
Similarly, in the work of~\citet{Palanisamy1}, a decentralized privacy-preserving transaction scheduling mechanism, enabling users to schedule transactions while safeguarding sensitive inputs until the execution time window specified by the users. 
The sensitive inputs are securely maintained by a randomly selected subset of trustees within the network, ensuring that the inputs are disclosed only at the designated execution time. In our solution, we employ attribute-based encryption to store only encrypted data on-chain without revealing it at any stage.
\citet{10.1145/3652162} present a private data marketplace where researchers and data owners can meet to find agreement on the use of private data for statistics or model trainings. In this solution Secure Multi-Party Computation (SMPC) is used to execute operations on private data, while Ethereum blockchain is used to create a campaign, to collect subscriptions and pay the services. In our approach, we employ multi-party computation in the system boot phase (\cref{sec:workflow:init}), while the smart contract deployed on the blockchain stores all the metadata and resource locators.\todo{ACM TOPS added}

\citet{chen2021blockchain} introduce a medical information system based on blockchain technology to safely store and share medical data. This work is based on Hyperledger Fabric~\citep{Baset.etal/2018:HyperledgerFabric}, a permissioned blockchain platform.
\citet{zhang2018towards} propose a blockchain-based secure and privacy-preserving personal health information sharing scheme, which resorts to private and consortium blockchains.
\citet{peng2023peer} propose the utilization of a consortium blockchain for the development of a decentralized storage system for file sharing between organizations. 
The data owner firstly encrypts the data and stores it on the consortium blockchain, and then authorizes the reader to access the data.
\citet{Cross-Organizational} 
describe a blockchain-based access control scheme to facilitate data sharing among different organizations.
They resort to a Role-Based Access Control (RBAC) model proposing a three-layered solution: A consortium blockchain at the network layer, dedicated smart contracts with a multi-signature processing node at the control layer, and the shared resources on the user layer.
The authors evaluate their approach using the HyperLedger Fabric consortium blockchain. 
\citet{xie2023tebds} present TEBDS, an IoT data sharing architecture employing Trusted Execution Environment (TEE)~\citep{TEE2,TEE1} and blockchain. TEBDS integrates both on-chain and off-chain approaches to fulfill the security needs of the IoT data sharing framework. 
The on-chain data security is achieved with a consortium blockchain while a TEE secures the off-chain IoT data.
The three-tier architecture Blockchain (TBchain)~\citep{TBchain}
is a solution that improves scalability and storage extensibility of blockchains to use them with IoT devices. The three tiers are the Super Blockchain (SB), the Middle Blockchain (MB) and the Underlying Blockchain (UB). TBchain is implemented as a second-layer private shard connected to Ethereum: 
The IoT devices save transaction data on TBchain, which 
collects this data into blocks and transfers their hash to Ethereum for validation and storage space optimization.
\citet{IoTDeviceBehavior}
introduce a privacy-preserving IoT device management framework based on blockchain technology. Within this framework, IoT devices are managed through multiple smart contracts deployed on a private Ethereum blockchain. These smart contracts authenticate the rules set by data owners, outlining permission configurations and recording any misconduct by individual IoT devices. Consequently, these smart contracts identify devices with vulnerabilities or those that have been compromised. As a result, data owner privacy is guaranteed by the monitoring activity and control conducted on the devices.
\citet{SharingSystems} and \citet{MedSBA} propose approaches for decentralized storage and sharing based on private blockchains. We operate in the context of public blockchains to leverage the higher degree of security given by the general validation of transactions.
Differently from all the approaches in this paragraph, MARTSIA is designed for ensuring data confidentiality with public permissionless blockchains.

\citet{nannawu} present an attribute-based access control scheme which grants the privacy of attributes and policies. Unlike MARTSIA, their solution requires a trusted entity that distributes keys and certificates to the other actors involved.
\citet{proud} introduce PROUD, a scheme for preserving privacy while outsourcing attribute-based signcryption. 
A trusted central authority is responsible for system initialization and users' private key creation, and a semi-trusted edge server is responsible for partial decryption of a ciphertext.
\citet{zyskind2015decentralizing} describe a decentralized system for managing personal data, empowering users with ownership and control over their data. In their work, the authors specifically concentrate on mobile platforms, where services are provided as user applications.
\citet{huang2020blockchain} present a blockchain-based privacy-preserving scheme that allows patients, research institutions, and semi-trusted cloud servers to share medical data. 
\citet{eltayieb2020blockchain} combine blockchain technology with attribute-based signcryption to provide a secure data sharing in the cloud environment. Their system ensures efficient access control over the data in the cloud server. To guarantee confidentiality, they resort to a Trust Authority that generates and distributes keys for data owners and readers.
\citet{BACKM-EHA}
propose BACKM-EHA, a blockchain-enabled access control and key management protocol for IoT-based healthcare applications. In their system, a trusted authority 
is necessary to register the communicating parties. Furthermore, personal servers and cloud servers are considered as the semi-trusted parties of the network.
\todo{Citation of ACM TOPS suggested editor: added}
\citet{BEAAS} introduce BEAAS+, an Attribute-Based Access Control (ABAC) as a service model built on an Ethereum-based protocol. BEAAS+ comprises three primary components: the Service Application (SA), the blockchain, and the Client Application (CA). The SA acts as the trusted entity in the system, incorporating a Verification Interface (VI) for resolving disputes through verification queries, an Application Server (AS) for handling service requests, and a Policy Store (PS), a back-end database that maintains the system state.
As opposed to the above approaches~\citep{nannawu,proud,zyskind2015decentralizing,huang2020blockchain,eltayieb2020blockchain,BACKM-EHA,BEAAS}, MARTSIA does not require the involvement of single trusted entities as its architecture is fully decentralized.

In the healthcare domain, \citet{EHR} create a secure electronic health records system that combines Attribute-Based Encryption, 
Identity-Based Encryption, 
and Identity-Based Signature 
with blockchain technology.
Their architecture is different from ours as the hospital has control over the patient's data and sets the policies, whereas solely the data owners manage data in MARTSIA.
\cite{zhang2018fhirchain} introduce FHIRChain, namely a blockchain-based architecture designed to share clinical data in compliance with the ONC requirements for IT in healthcare. 
To protect the content they use a secure cryptographic mechanism called \textit{sign then encrypt}. MARTSIA, in contrast, uses \MAABE to encrypt the data.
\citet{IoT-ABE} propose a new blockchain architecture for IoT applications that preserves privacy through the use of \ABE. We also utilize a similar kind of encryption in our approach, but
MARTSIA integrates with existing technologies, whilst their model aims to change the blockchain protocol.
\citet{B-Box} propose an idea for a decentralized storage system named B-Box, based on \IPFS, \MAABE and blockchains. Though we resort to those building blocks too, we include mechanisms for secure initialization of the authority network, allow for fine-grained access control on message parts, and impede by design any actor from accessing keys.

\subsection*{Role-based access control blockchain-based approaches leveraging \acrlong{abe}}\label{sec:sota:abe}~\citet{Linjian} propose a secure and efficient data-sharing solution based on decentralized ABE (\MAABE), blockchain, and IPFS. Unlike our solution, their architecture needs two trusted parties, namely the key generation centre (KGC) and the Attribute Authority (AA), to generate the public parameters. In our work, all the authorities participate in a simple commit-then-open coin-tossing protocol to create such parameters when the MA-ABE is instantiated considering the work of~\citet{Rouselakis}. Moreover, we prove the security of our solution in the Universal Composability (UC) framework~\cite{Canetti01}. 
\citet{secrypt18} present BDABE, an access control mechanism that applies a distributed ABE scheme to the real world using a consensus-driven infrastructure. In their solution, a semi-trusted Attribute Authority (AU) is responsible for creating users and assigning attributes to them. A central, fully trusted Root Authority (RA) is responsible for making the secret keys of AUs and handing them to an AU. 
In contrast, we do not employ semi or fully-trusted authorities since any protocol phase, including attribute key generation, is decentralized.
\citet{Healthcare} introduce BABEHealth, a system that improves the security and privacy of exchanging electronic healthcare records using blockchain technology, IPFS, and attribute-based encryption. Unlike our architecture, they employ a private blockchain and do not use decentralized ABE. 
As in the work of~\citet{secrypt18} the setup and trusted authorities are not decentralized, and no formal security proof is reported. 
\citet{info14050281} propose a scheme combining attribute-based encryption ABE and identity-based encryption (IBE) to achieve efficient data sharing and verification of data correctness. They also employ blockchain technology to ensure tamper-proof and regulated data storage. In contrast with our architecture, they do not employ decentralized ABE. Also, the setup phase and the attribute key generation are centralized because a given Private Key Generator is in charge.
\citet{Wu2019} present an efficient and privacy-preserving attribute-based encryption scheme that uses blockchains to guarantee both integrity and non-repudiation of data. In their solution, the version of ABE in use is not decentralized, and the public parameters are generated and published by an Attribute Authority (AA). On the contrary, our solution uses decentralized attribute-based encryption, and the public parameters are generated via MPC rather than by a single, fully-trusted entity.
\citet{PRShare} extends ABE with a novel technique called Attribute-Based Encryption with Oblivious Attribute Translation (OTABE). OTABE is the crucial part of their architecture that facilitate inter-organizational data sharing. This extension enhances ABE by supporting dynamic and oblivious translation of proprietary attributes across organizations while maintaining hidden access policies, direct revocation capabilities, and fine-grained, data-centric key and query functionalities.
\todo{ACM TOPS added}
\citet{Yan2023} introduce a scheme that allows for fine-grained access control with low computation consumption by implementing proxy encryption and decryption while supporting policy hiding and attribute revocation. Similarly to our solution, the encrypted data is stored on IPFS, and the metadata is stored on the blockchain. However, their work does not employ decentralized ABE, as a centralized authority generates the setup and the necessary keys. Also, two centralized entities are involved, namely the Proxy encryption server (ES) and the Proxy decryption server (DS), that are responsible for proxy encryption and decryption calculation, respectively.
	

%% file: sections/conclusion.tex
In this work, we described MARTSIA, a technique leveraging \acrfull{maabe} to regulate data access in multi-party business operations supported by blockchain technology. Our approach also employs \IPFS to preserve information artifacts, access regulations, and metadata.
We utilize smart contracts to keep the user attributes, establish the access grants of the process participants, and save the locators of IPFS files. MARTSIA allows a highly detailed specification of access permissions, ensuring data reliability, persistence, and irrefutability in a completely decentralized setting. Thus, it enables auditability with minimal added costs.
We discuss and prove the security guarantees our solution carries from a formal standpoint.
MARTSIA is platform-independent, as our proof-of-concept implementations for \EVM and \AVM-based platforms demonstrate.
Also, it is interoperable, as the plug-ins we created for interfacing our solution to existing tools and applications show, introducing limited overhead in terms of cost and execution time.

Our approach exhibits limitations we aim to overcome in future work. If a \DOwner wants to revoke access to data for a particular \Reader, e.g., they can change the policy and encrypt the messages again. However, the old data on \IPFS would still be accessible. Therefore, we are considering the usage of InterPlanetary Name System (IPNS), as it allows for the replacement of existing files. With it, a message can be replaced with a new encryption thereof that impedes {\Reader}s whose grant was revoked to access it.
More generally, the life-cycle of data artifacts, policies and smart contracts constitutes a management aspect worth investigating.
In light of the considerable impact that a correct expression of policies has on the overall approach, we envision automated verification and simulation of policies for future work to properly assist the users tasked with their specification.
Future endeavors also include the integration of Zero Knowledge Proofs~\citep{goldwasser2019knowledge} with \ABE to yield better confidentiality and privacy guarantees, and of decentralized identifiers~\citep{Norta.etal/CS2019:BlockchainEnabledIdentityAuth} and oracles~\citep{Basile.etal/BPMBCF2021:BlockchainProcessesDecentralizedOracles} 
to verify data ownership.
We plan to 
run field tests for the empirical evaluation of the robustness of our approach, possibly integrating MARTSIA with more blockchain-based data and workflow management tools such as ChorChain~\citep{Corradini.etal/ACMTMIS2022:EngineeringChoreographyBlockchain} and FlexChain~\cite{Corradini.etal/FGCS2023:FlexChain}. 
An interesting research direction is the construction of solutions not based on ABE. In this context, the idea is to build MPC-based protocols inspired by~\citet{Choudhuri.etal/CRYPTO2021:FluidMPC} and \citet{Gentry.etal/CRYPTO2021:YOSO} that are tailored, efficient and publicly verifiable with the aim of achieving private data retrieval.
Finally, we are currently working on a solution that uses Trusted Execution Environments (TEEs)~\citep{TEEfirst,TEEsecond} for the exposition of aggregated and manipulated data to aid automatic decision-making processes in a workflow management tool.